\documentclass[aps,prl,twocolumn,amsmath,amssymb,10pt,aps,longbibliography,superscriptaddress,citeautoscript,bibnotes,floatfix,nobibnotes]{revtex4-2}

\usepackage[utf8]{inputenc}
\usepackage[T1]{fontenc}
\usepackage{textcomp} %% provides additional text symbols

\usepackage{amsmath}   %% basic math formatting
\usepackage{amssymb}   %% this does nothing without amsfonts?
\usepackage{physics}   %% various physics notation, e.g. bra-ket
\usepackage{bm}        %% bold math symbols, e.g. greek letters which do not have a \mathbf
\usepackage{mathrsfs}  %% \mathsf font
\usepackage{mathtools} %% additional features to amsmath, e.g. dcases environment
\newcommand*{\diff}{\mathop{}\!\mathrm{d}} %% differential operator
\newcommand*{\Diff}{\mathop{}\!\mathrm{D}} %% differential operator

\usepackage{caption}  %% additional control of the captions in floating environments
\usepackage{graphicx} %% additional control of \includegraphics, extension of {graphics}
\graphicspath{ {./Figures/} }

\usepackage{subfigure} %% deprecated, should be replaced with {subcaption}

\usepackage{tabularx} %% extension of {tabular} with adjustable-width columns
\usepackage{booktabs} %% internal optimizations and additional control of table environments
\usepackage{multirow} %% create tabular cells spanning multiple rows
\usepackage{dcolumn}  %% align table columns on decimal point

\usepackage[colorlinks=true,linkcolor=blue,citecolor=blue]{hyperref} 

\usepackage[normalem]{ulem} %% provides underline command, \ul
\usepackage{soul} %% spacing out (letterspacing), underlining, striking out, etc.
\usepackage{amsthm}  %% AMS formatting of {theorem} environments
\usepackage{color,xcolor} %% foreground (text, rules, etc.) and background colour management
\usepackage{lipsum}

\usepackage{titlesec}
\titleformat{\section}{\normalfont\large\bfseries}{}{}{}
\titlespacing{\section}{0pt}{10pt plus 2pt minus 2pt}{0pt}

\titleformat{\paragraph}{\normalfont\bfseries}{}{}{}
\titlespacing{\paragraph}{0pt}{10pt plus 2pt minus 2pt}{0pt}

\usepackage{comment}
\usepackage{soul}

\begin{document}
\title{Superconductivity Proximate to Non-Abelian Fractional Spin Hall Insulator in Twisted Bilayer MoTe$_2$}

\author{Cheong-Eung Ahn}
\altaffiliation{These authors contributed equally to this work.}
\affiliation{Department of Physics, Pohang University of Science and Technology, Pohang, 37673, Republic of Korea}

\author{Donghae Seo}
\altaffiliation{These authors contributed equally to this work.}
\affiliation{Department of Physics, Pohang University of Science and Technology, Pohang, 37673, Republic of Korea}

\author{Gyeoul Lee}
\affiliation{Department of Physics, Ulsan National Institute of Science and Technology, Ulsan 44919, Republic of Korea}

\author{Youngwook Kim}
\affiliation{Department of Physics and Chemistry, Daegu Gyeongbuk Institute of Science and Technology (DGIST), Daegu 42988, Republic of Korea}

\author{Gil Young Cho}
\thanks{gilyoungcho@kaist.ac.kr}
\affiliation{Department of Physics, Korea Advanced Institute of Science and Technology, Daejeon 34141, Republic of Korea}
\affiliation{Center for Artificial Low Dimensional Electronic Systems, Institute for Basic Science, Pohang 37673, Korea}

\begin{abstract}
Twisted bilayer MoTe$_2$ near two-degree twists has emerged as a platform for exotic correlated topological phases, including ferromagnetism and a non-Abelian fractional spin Hall insulator. Here we reveal the unexpected emergence of an intervalley superconducting phase that intervenes between these two states in the half-filled second moir\'{e} bands. Using a continuum model and exact diagonalization, we identify superconductivity through multiple signatures: negative binding energy, a dominant pair-density eigenvalue, finite superfluid stiffness, and pairing symmetry consistent with a time-reversal-symmetric nodal extended $s$-wave state. Remarkably, our numerical calculation suggests a continuous transition between superconductivity and the non-Abelian fractional spin Hall insulator, in which topological order and symmetry evolve simultaneously, supported by an effective field-theory description. Notably, our field-theoretic analysis indicates that superconductivity is driven by the condensation of charge-$e/2$ self-bosonic non-Abelian anyons, thereby providing a concrete realization of anyon superconductivity. Complementarily, when approached from the normal metallic side, superconductivity instead emerges from a Kohn–Luttinger instability enabled by the non-uniform quantum geometry of the flat moiré bands. Our results establish higher moir\'{e} bands as fertile ground for intertwined superconductivity and topological order, and point to experimentally accessible routes for realizing superconductivity in twisted bilayer MoTe$_2$.
\end{abstract}

\maketitle

\section{Introduction} \noindent
Flat topological bands with strong electron correlations have recently attracted significant attention for realizing long-sought topological phases~\cite{cao2018unconventional,chen2019signatures,cao2018correlated,regan2020mott,cao2020strange}, such as fractional Chern insulators~\cite{xu2023observation,cai2023signatures,zeng2023thermodynamic,park2023observation,ji2024local,redekop2024direct,park2025observation,kang2024evidence,kang2025time,lu2024fractional,lu2025extended,xie2025tunable,aronson2025displacement}, which are promising candidates for topological quantum computation~\cite{nayak2008non}. A particularly intriguing platform is small-angle twisted bilayer MoTe$_2$, where the second moir\'{e} bands are predicted to host non-Abelian fractional Chern insulators~\cite{ahn2024non,reddy2024non,chen2025robust,xu2025multiple,wang2025higher,liu2025non}, while experiments have uncovered an even more exotic phase—a non-Abelian fractional spin Hall insulator~\cite{kang2024evidence,kang2025time}. Notably, in these flat Chern band systems, chiral superconductivity has also been experimentally observed~\cite{xu2025signatures}, generating considerable interest in uncovering the relationship between superconductivity and topological phases~\cite{guerci2025fractionalization,xu2025chiral,shi2025doping,shi2025doping_,pichler2025microscopic,wang2025chiral,zhang2025holon,nosov2025anyon,shi2025anyon,huang2025apparent}.

In this manuscript, we identify another intriguing quantum state, namely an intervalley superconducting phase, that can arise in this strongly correlated topological flat band system. Specifically, we report evidence for such a phase proximate in the phase diagram to the non-Abelian fractional spin Hall insulator in the half-filled second moir\'{e} bands, i.e., hole filling $\nu_h =3$, of approximately 2$^{\circ}$-twisted bilayer MoTe$_2$~\cite{kang2024evidence,kang2025time,abouelkomsan2025non}. We perform exact diagonalization on a continuum model of the half-filled second moir\'{e} bands, incorporating both valleys and screened Coulomb interactions~\cite{abouelkomsan2025non}. Superconductivity emerges when the short-range component of the intervalley interactions is screened, yielding time-reversal symmetry and extended $s$-wave pairing.

We further propose two complementary pairing mechanisms underlying the observed superconductivity: a Kohn–Luttinger pairing instability~\cite{kohn1965new} on the metallic side and anyon superconductivity~\cite{laughlin1988superconducting,laughlin1988relationship} emerging from the non-Abelian spin Hall insulator. The former originates from the screening of repulsive Coulomb interactions, which generates an effective attractive pairing channel, similar to recent studies of superconductivity in the first moir\'{e} band of twisted bilayer MoTe${}_2$~\cite{guerci2025fractionalization,xu2025chiral}. Crucially, we find that the non-uniform quantum geometry of the moir\'e bands plays an essential role in enabling intervalley pairing. For example, when the second moir\'e bands in our model are replaced by first Landau levels with uniform quantum geometry, the superconducting tendency is strongly suppressed~\cite{abouelkomsan2025non}. This comparison underscores the importance of realistic band geometry in stabilizing superconductivity, in agreement with recent theoretical developments~\cite{guerci2025fractionalization,xu2025chiral,shi2025doping,shi2025doping_,pichler2025microscopic,wang2025chiral,zhang2025holon,nosov2025anyon,shi2025anyon,huang2025apparent,shavit2025quantum,jahin2026enhanced}.

Remarkably, our numerics uncovers a continuous single transition from the fractional spin Hall insulator to the superconducting state, with spectral flows closely resembling those of the continuous transition from a bilayer fractional quantum Hall state composed of two Pfaffian copies to an exciton condensate~\cite{shi2008phase,zhu2019exciton}. This transition is further supported by our effective field-theory description, where superconductivity emerges from the fractional spin Hall insulator via the condensation of non-Abelian charge-$e/2$ anyons. In this description, the transition is unconventional and and lies beyond the traditional Landau–Ginzburg paradigm: it simultaneously alters the global charge-conservation symmetry and the underlying topological order. At the same time, it provides a pairing mechanism in which superconductivity arises from the condensation of charged anyons rather than a Cooper pair, a mechanism that has recently attracted significant attention~\cite{shi2025doping,shi2025doping_,pichler2025microscopic,wang2025chiral,zhang2025holon,nosov2025anyon,shi2025anyon,huang2025apparent,zhang2025charge4e,kuhlenkamp2025robust,shi2026charge4e,seo2026unified}.

In summary, our results in this manuscript suggest that higher moir\'{e} bands may provide a promising platform to explore the interplay between superconductivity, strong correlations, quantum geometry, and topological orders, thereby broadening the landscape of superconductivity and fractionalization beyond previous studies~\cite{shi2025doping,shi2025doping_,pichler2025microscopic,wang2025chiral,zhang2025holon,nosov2025anyon,shi2025anyon,huang2025apparent,zhang2025charge4e,kuhlenkamp2025robust,shi2026charge4e,seo2026unified,laughlin1988superconducting,laughlin1988relationship}.

\begin{figure*}[t]
\centering
\includegraphics[width=0.99\textwidth]{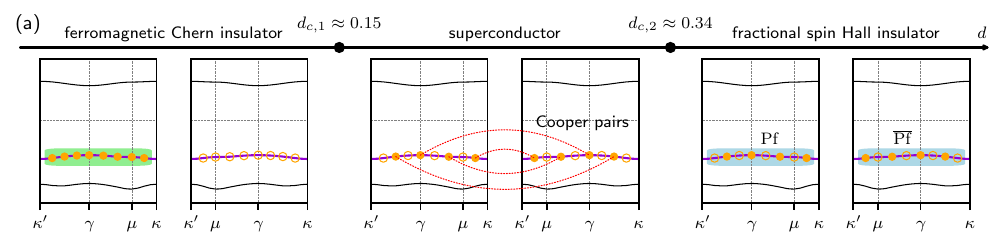}
\includegraphics[width=0.75\textwidth]{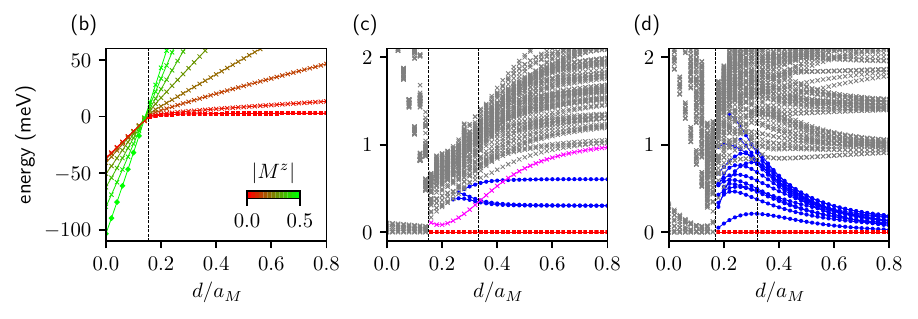}
\caption{\textbf{Schematic phase diagram and evolution of many-body spectrum.} (a) Schematic phase diagram as a function of $d$ in Eq.~\eqref{eq:interaction}, with $a_M = 1$. As $d$ increases from $0$ to $+\infty$, the system evolves through a ferromagnetic Chern insulator, a superconductor, and finally a fractional spin Hall insulator. (b) The lowest energy of each magnetization subspace for $N=14$. (c,d) Evolution of the many-body spectrum within the paramagnetic sector $M_z =0$ for $N=14$ and 12, respectively. The evolution of the many-body spectrum suggests there are two transition points, a first-order transition $d_{c,1}$ which changes the magnetization, and another continuous transition at $d_{c,2}$ which changes the nature of the ground-state manifold and excitations.}
\label{fig:M1_PhaseDiagram}
\end{figure*}

\section{Results}
\paragraph{The Model.}\noindent
As an effective description of twisted bilayer MoTe${}_2$, we adopt the continuum model~\cite{wu2019topological}. The non-interacting Hamiltonian is decomposed into the $K$- and $K'$-valleys, $\hat{h}=\hat{h}_\uparrow \oplus \hat{h}_\downarrow$, with the $K$-valley Hamiltonian given by
\begin{equation}
\hat{h}_\uparrow=\begin{bmatrix}
-\frac{\left(\hat{\mathbf{k}}-K_+\right)^2}{2m^*}+v_+\left(\hat{\mathbf{r}}\right) & \gamma^*\left(\hat{\mathbf{r}}\right)\\
\gamma\left(\hat{\mathbf{r}}\right) & -\frac{\left(\hat{\mathbf{k}}-K_-\right)^2}{2m^*}+v_-\left(\hat{\mathbf{r}}\right)
\end{bmatrix},\label{eq:single-electron-hamiltonian-K}
\end{equation}
where $\hat{\mathbf{k}}$ denotes the electron momentum, $K_\pm = R_{\theta/2} K$ are the twisted $K$ points of the two layers by angle $\theta$, and $v_\pm(\mathbf{r})$ and $\gamma(\mathbf{r})$ represent the moir\'{e} potentials and interlayer tunneling, respectively. Note that the spin and valley are locked due to the Ising spin-orbit coupling~\cite{xiao2012coupled}. In this manuscript, we focus primarily on the system with a twist angle $\theta$ close to 2$^{\circ}$. The $K'$-valley Hamiltonian is obtained as the time-reversal conjugate of Eq.~\eqref{eq:single-electron-hamiltonian-K}, ensuring that the entire system is time-reversal symmetric. The explicit forms of $v_\pm(\mathbf{r})$ and $\gamma(\mathbf{r})$ are specified in~\cite{supp}.

We also include the following interaction in our model,
\begin{equation*}
\hat{H}=-\hat{h}+\frac{1}{2A}\sum_{s_0,s_1\in\{\uparrow,\downarrow\}, \mathbf{p}} V_{s_0s_1}\left(\mathbf{p}\right):\hat{\rho}^{s_1}_{-\mathbf{p}}\hat{\rho}^{s_0}_{\mathbf{p}}:.
\end{equation*}
Here, $\hat{\rho}^s_{\mathbf{p}}$ denotes the electron density operator of momentum $\mathbf{p}$ with spin $s \in \{\uparrow,\downarrow\}$, and {$A$ is the sample area}. The interaction $V_{s_0s_1}\left(\mathbf{p}\right)$ is given by
{
\begin{equation}
V_{s_0s_1}\left(\mathbf{p}\right)=\frac{e^2}{2\epsilon}\begin{dcases}
\frac{1}{p} & s_0=s_1\\
\frac{e^{-pd}}{p} & s_0\neq s_1
\end{dcases}.\label{eq:interaction}
\end{equation}}\noindent
Notably, while the intravalley interaction ($s_0 = s_1$) is simply given by the Coulomb potential, the intervalley interaction ($s_0 \neq s_1$) is the screened form arising from band mixing {with an effective screening length $d$}, as phenomenologically introduced in \cite{abouelkomsan2025non}.

A few remarks are in order. First, at a twist angle near 2$^{\circ}$, the $K$-valley of twisted bilayer MoTe${}_2$ realizes a sequence of Chern bands, all with $C=+1$, closely resembling the sequential Landau levels of a conventional two-dimensional electron gas~\cite{shi2024adiabatic}. The resemblance is not limited to the band topology alone: the effective interactions within each moir\'{e} bands strikingly mirror the Haldane pseudopotentials of their corresponding Landau levels~\cite{ahn2024non}. Secondly, the moir\'{e} bands possess almost ideal dispersion and geometry~\cite{ahn2024non,reddy2024non,chen2025robust,xu2025multiple,wang2025higher,liu2025non}, such that a half-filled \textit{single-valley} second moir\'{e} band—remarkably akin to the first Landau level—can stabilize a non-Abelian fractional Chern insulator~\cite{ahn2024non,reddy2024non,chen2025robust,xu2025multiple,wang2025higher,liu2025non}, specifically the Pfaffian state~\cite{reddy2024non}. This may be relevant to the resistivity plateau near the filling $\nu_h = 3/2$ observed in experiment~\cite{kang2024evidence,kang2025time}. The band dispersion and quantum geometry of our model can be found in~\cite{supp}.

\begin{figure}[b!]
\centering
\includegraphics[width=0.99\linewidth]{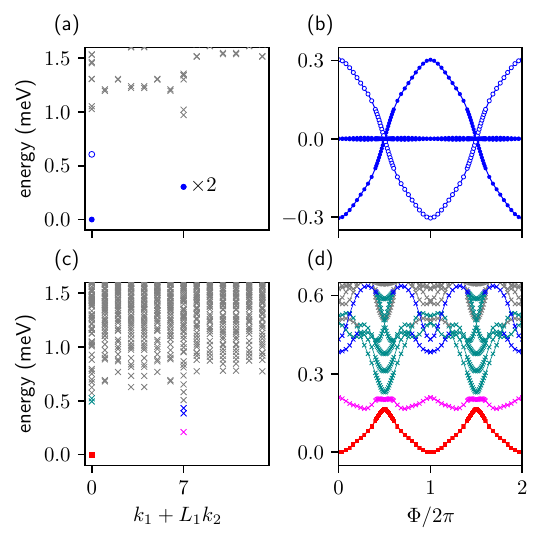}
\caption{\textbf{Many-body spectrum and spectral flow for $N=14$.} (a) Many-body spectrum at $d/a_M=0.80$ (blue circle and gray x markers). (b) Spectral flow under time-reversal symmetric flux insertion at $d/a_M=0.80$. We find a $4\pi$-periodic spectral flow which mixes the four lowest states (blue circles), consistent with the fractional spin Hall insulator. (c) Many-body spectrum at $d/a_M=0.28$ (red square, magenta, blue, cyan, and gray x markers). The red square (ground-state) and two blue x markers are the wavefunctions with greatest similarity to three of the four ground-state wavefunctions of the fractional spin Hall insulator. The two cyan x markers both have large contributions from the last fractional spin Hall insulator ground-state wavefunction. The first excited state is the magenta x marked state, which is not adiabatically connected to the fractional spin Hall insulator ground-state wavefunctions. (d) Under time-reversal symmetric flux insertion of (c), the unique ground-state at $d/a_M=0.28$ remains isolated from the excited states. The spectral flow is $2\pi$-periodic.}
\label{fig:M2_ManyBodySpectrum}
\end{figure}

\paragraph{Phase Diagram.}\noindent
We map out the phase diagram [Fig.~\ref{fig:M1_PhaseDiagram}(a)] of the half-filled second moir\'{e} bands, i.e., hole filling $\nu_h = 3$, as $d$ in Eq.~\eqref{eq:interaction} is tuned, which hosts both ferromagnetic Chern insulators and fractional spin Hall insulators, as well as a possible intermediate phase. Notably, both ferromagnetism and fractional spin Hall insulators have already been observed experimentally in different devices at the corresponding filling near 2$^{\circ}$ twist~\cite{kang2024evidence,kang2025time,li2025universal}.

To build intuition, we first examine the two limiting cases of Eq.~\eqref{eq:interaction}, $d \to 0$ and $d \to +\infty$. In the $d \to 0$ limit, intervalley and intravalley interactions become indistinguishable, and Stoner’s mechanism drives ferromagnetism. The holes then spontaneously choose one valley to occupy, giving rise to magnetization—or equivalently, valley polarization. Because the occupied band carries a Chern number, this realizes a ferromagnetic Chern insulator. At the opposite limit, $d \to +\infty$, the valleys decouple entirely. Each half-filled moir\'{e} band, forming a time-reversal pair, stabilizes the Pfaffian state and its time-reversal conjugate, together yielding the fractional spin Hall insulator~\cite{abouelkomsan2025non}. As both are gapped phases, their stability is expected to extend to finite $d$~[Fig.~\ref{fig:M1_PhaseDiagram}(a)]. 

We employ exact diagonalization (ED) to systematically examine the stability of both phases with varying $d$. ED is performed with periodic boundary conditions on system sizes $N=12, 14,$ and $16$~\cite{supp}. Our ED results largely confirm the intuition developed above, but also reveal an unexpected intermediate phase~[Fig.~\ref{fig:M1_PhaseDiagram}(a)]. For example, the ground-state energies in each magnetization sector~[Fig.~\ref{fig:M1_PhaseDiagram}(b)] show that the model stabilizes a ferromagnetic Chern insulator for $d < d_{c,1} \approx 0.15\,a_M$ as expected, while it becomes paramagnetic for $d > d_{c,1}$. Similarly, we observe clear stabilization of the fractional spin Hall insulator for $d > d_{c,2} \approx 0.34\,a_M$~[Fig.~\ref{fig:M1_PhaseDiagram}(c,d)], where the system shows a ground-state degeneracy of 36 for $N = 12,16$~\cite{supp} and 4 for $N = 14$~[Fig.~\ref{fig:M2_ManyBodySpectrum}(a)], consistent with the even–odd effect expected for this phase~\cite{oshikawa2007topological}. The $4\pi$-periodic spectral flow of the ground-states in this regime under time-reversal symmetric flux insertion, confirmed by tracking the wavefunction fidelities, further supports this~[Fig.~\ref{fig:M2_ManyBodySpectrum}(b)].
 
Beyond the two anticipated phases, we uncover an intermediate phase, characterized by a singly-degenerate paramagnetic ground-state appearing for $d \in (d_{c,1}, d_{c,2})$~[Fig.~\ref{fig:M2_ManyBodySpectrum}(c)]. Tracking the wavefunction fidelities, we find that the other low-lying states that previously belonged to the ground-state manifold of the fractional spin Hall insulator now hybridize with excited states. Furthermore, the change in ground-state degeneracy compared to the fractional spin Hall insulator ($d > d_{c,2}$) is not merely a finite-size effect, as evidenced by the distinct evolution of the many-body spectrum under time-reversal symmetric flux insertion. Unlike the $4\pi$-periodic spectral flow of the fractional spin Hall insulator, the unique ground-state of the intermediate phase remains isolated from excited states and exhibits a $2\pi$-periodic spectral flow~[Fig.~\ref{fig:M2_ManyBodySpectrum}(d)]. The change in many-body topology is further evidenced by the spin Hall conductivity which changes from $\sigma_s = e^2/h$ for $d > d_{c,2}$ to $\sigma_s = 0$ for $d_{c,1} < d < d_{c,2}$~\cite{supp}. Finally, we also find that the ground-state energy derivatives and ground-state fidelity indicate a continuous phase transition at $d = d_{c,2}$ between the intermediate phase and the fractional spin Hall insulator~\cite{supp}.

Both the features and spectral flow of the ground-states closely resemble those of the continuous transition from bilayer fractional quantum Hall states of two Pfaffian copies to an exciton condensate~\cite{shi2008phase,zhu2019exciton}. As we will show later, this continuous transition is consistent with our identification of the intermediate state as a superconducting phase. In \cite{supp}, we further present data demonstrating the enhanced transition behavior when the short-range repulsion—specifically, the zeroth Haldane pseudopotential~\cite{kwan2026regarding}—is explicitly suppressed.

In~\cite{supp}, we also present the many-body spectra and their evolution in the generalized first Landau level and its time-reversal conjugate with non-uniform quantum geometry. In the idealized first Landau level at $N=14$ with uniform quantum geometry, the system appears to undergo a direct transition from the ferromagnetic Chern insulator to the fractional spin Hall insulator, with no intermediate phase showing a 1-fold degenerate ground-state~\cite{abouelkomsan2025non}. However, upon introducing non-uniform quantum geometry~\cite{wang2021exact,fujimoto2025higher,liu2025theory}, the intermediate state emerges and becomes visible. This demonstrates that non-uniform quantum geometry plays a crucial role in stabilizing the intermediate phase, even without explicit band dispersion.

\begin{figure}[b!]
\centering
\includegraphics[width=0.99\linewidth]{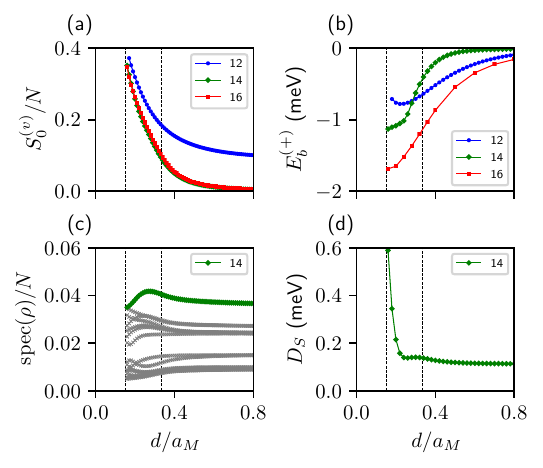}
\caption{\textbf{Signatures of superconductivity.} (a) Intervalley entanglement entropy $S_0^{(v)}$, (b) Binding energy $E_b^{(+)}$, (c) Spectrum of the normalized pair density matrix $\rho_{\mathbf{k}'\mathbf{k}}/N$, and (d) Superfluid stiffness $D_S$ of the ground-state over $d$.}
\label{fig:M3_SCSignatures}
\end{figure}

\paragraph{Intermediate Superconductivity.}\noindent
We now present several pieces of evidence indicating that the intermediate state is a superconductor. Our first clue from the many-body spectrum is that the ground-state carries zero total magnetization and momentum across all numerically-accessible system sizes and neighboring fillings in the intermediate regime $d\in(d_{c,1},d_{c,2})$~\cite{supp}. This is consistent with Cooper pairing between holes of opposite spins and valleys at opposite momenta, i.e., between $\left(\uparrow,\mathbf{k}\right)$ and $\left(\downarrow,-\mathbf{k}\right)$. An additional signature is seen in the intervalley entanglement entropy, $S^{(v)}_0=-\trace\left(\hat{\rho}_{0,\uparrow}\ln\hat{\rho}_{0,\uparrow}\right)$, which changes abruptly from large values in the intermediate regime $d \in (d_{c,1}, d_{c,2})$ to small values in the fractional spin Hall insulators ($d > d_{c,2}$)~[Fig.~\ref{fig:M3_SCSignatures}(a)]. Here $\hat{\rho}_{0,\uparrow}$ is the $K$-valley reduced density operator of the ground-state wavefunction, obtained by partially tracing out the $K'$-valley Hilbert space. This behavior further supports the picture of intervalley pairing, which inherently links and entangles the two valleys~[Fig.~\ref{fig:M1_PhaseDiagram}(a)].

We further observe that the intermediate state exhibits a quantitatively enhanced tendency to form Cooper pairs, as indicated by the binding energy,
\begin{equation*}
E_b^{(\pm)} = E_{N\pm 2,0} + E_{N,0} - 2E_{N\pm 1,0},
\end{equation*}
where $E_{N_h,0}$ is the ground-state energy for $N_h$ holes. Across all three system sizes $N=12, 14, 16$, we find that the binding energy is negative for all $d$, with its magnitude $|E_b|$ comparable to the Fermi energy $E_F\approx 1.50\:\mathrm{meV}$ for the intermediate regime $d\in\left(d_{c,1},d_{c,2}\right)$, significantly smaller for $d>d_{c,2}$, and attaining a maximum slope at $d\approx d_{c,2}$~[Fig.~\ref{fig:M3_SCSignatures}(b)]. The identification of the intermediate state as a superconductor is further corroborated by the spectrum of the pair density matrix~\cite{yang1962concept},
\begin{equation}
\rho_{\mathbf{k}'\mathbf{k}}=\ev{\hat{\Delta}_{\mathbf{k}'}^\dagger\hat{\Delta}_{\mathbf{k}}}_0,\qquad\hat{\Delta}_{\mathbf{k}}^\dagger=\hat{\psi}_{\uparrow,\mathbf{k}}^\dagger\hat{\psi}_{\downarrow,-\mathbf{k}}^\dagger,
\label{eq:pair-density-matrix}
\end{equation}
where $\hat{\psi}_{s,\mathbf{k}}^\dagger$ is the operator that creates a hole in the second moir\'{e} band with spin $s$ and crystal momentum $\mathbf{k}$. We observe that the pair density matrix shows a single dominant eigenvalue, well separated from the rest of the spectrum, and also this eigenvalue reaches its maximum within the intermediate regime $d \in (d_{c,1}, d_{c,2})$~[Fig.~\ref{fig:M3_SCSignatures}(c)]. Both results show a tendency for binding, specifically between $\left(\uparrow,\mathbf{k}\right)$ and $\left(\downarrow,-\mathbf{k}\right)$, which is maximized in the intermediate region. In \cite{supp}, we make a similar analysis of the excitonic pair density matrix, and rule out intervalley exciton condensation as a candidate for the intermediate phase.

A defining hallmark of superconductivity is global phase coherence, captured by the superfluid stiffness $D_S$. We evaluate $D_S$ from the curvature of the ground-state energy with respect to global flux insertion around the non-contractible loops of the torus~\cite{scalapino1993insulator,resta2018drude}. We find that the superfluid stiffness reaches a maximum at $d_{c,1}$ and falls to zero as $d$ approaches $d_{c,2}$~[Fig.~\ref{fig:M3_SCSignatures}(d)]. This indicates that the intermediate region maximizes not only the pairing tendency but also phase coherence, establishing it as a superconductor.

\begin{figure}[b!]
\centering
\includegraphics[width=0.99\linewidth]{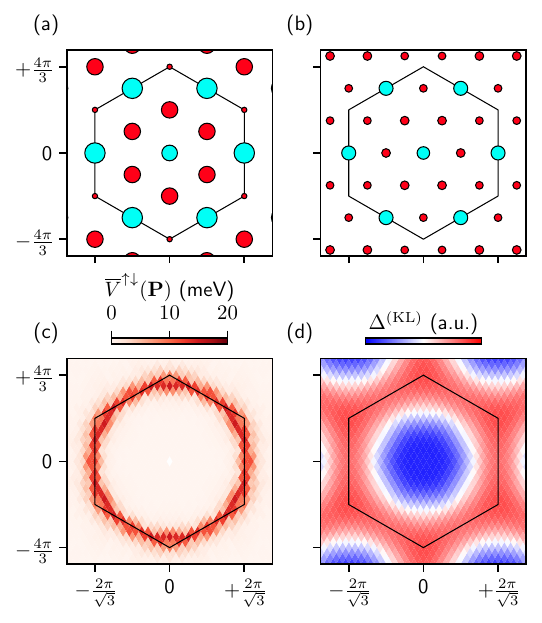}
\caption{\textbf{Pairing symmetry and Kohn-Luttinger instability.} (a,b) Cooper-pair order parameter for $N=12$ and $16$, respectively. The size of the circle is {proportional} to the absolute value of the order parameter at the corresponding momenta, and red and cyan colors indicate opposite signs. (c) Random-phase-approximation-screened intervalley interaction. (d) Corresponding leading pairing instability. All calculations are done for $d/a_M=0.28$.}
\label{fig:M4_PairingSymmetry}
\end{figure}

Building on these results, we can now analyze the pairing symmetry of the superconducting state by evaluating the order parameter
\begin{equation*}
\Delta_{\mathbf{k}}=\mel{N,0}{\hat{\psi}_{\downarrow,-\mathbf{k}}\hat{\psi}_{\uparrow,\mathbf{k}}}{N+2,0},
\end{equation*}
with $\ket{N_h,0}$ denoting the many-body ground-state for $N_h$ holes. The resulting wavefunction~[Fig.~\ref{fig:M4_PairingSymmetry}(a,b)] is real, ensuring time-reversal symmetry, and its momentum-space structure belongs to the trivial irreducible representation ($A_1$) of the point group symmetry generated by $C_3^z$ and $C_2^y\mathcal{T}$. The sign change across the Brillouin zone naturally classifies the superconducting state as a nodal extended $s$-wave superconductor. As we note in~\cite{supp}, the wavefunction that maximizes the pair density matrix Eq.~\eqref{eq:pair-density-matrix} exhibits the same pairing symmetry, providing further confirmation of the result. The nodal character of the resulting superconductivity is further corroborated by the observation that the superfluid stiffness $D_s$ is comparable to the binding energy $|E_b|$, reaching a maximum ratio of $D_s/(|E_b|/2) \approx 0.9$. This stands in contrast to conventional fully gapped superconductors, where $D_s \gg |E_b|/2$~\cite{emery1995importance,chen2024superconductivity}.

In~\cite{supp}, we report several additional observations. First, as expected, we observe that the many-body spectrum in our model is adiabatically connected to that of an ideal extended $s$-wave superconductor, realized when a strong attractive interaction in this pairing channel is introduced. In addition, we have also computed the charge gap, whose behavior is again consistent with a nodal superconductor. Finally, we find that the binding energy and Fermi energy are of comparable magnitude and estimate a relatively short superconducting coherence length~\cite{supp}, placing the intermediate superconducting state in the BCS--BEC crossover regime~\cite{chen2024superconductivity}, consistent with experimental observations in moir\'{e} flat band superconductors of magic-angle twisted bilayer graphene~\cite{oh2021evidence,tian2023evidence}, magic-angle twisted trilayer graphene~\cite{park2021tunable,kim2022evidence}, and first moir\'{e} band twisted bilayer MoTe${}_2$~\cite{xu2025signatures}. We note that our estimate for superconducting coherence length incorporates quantum geometric contributions~\cite{iskin2023extracting,chen2024ginzburg,hu2025anomalous,iskin2025pair,thumin2025correlation}.

\paragraph{Kohn-Luttinger Instability.}\noindent
The superconductivity observed in our exact diagonalization calculations can be understood in terms of the Kohn--Luttinger mechanism~\cite{kohn1965new}. In particular, the continuum moir\'e band model in Eq.~\eqref{eq:single-electron-hamiltonian-K} exhibits both finite dispersion and non-uniform quantum geometry, the latter effectively contributing to dispersion effects when expressed in the hole representation~\cite{guerci2025fractionalization,supp}. As detailed in~\cite{supp}, we find that several factors---the presence of multiple Fermi surface pockets~\cite{maiti2010renormalization}, the proximity of van Hove singularities to the Fermi surfaces~\cite{gonzalez2008kohn,nandkishore2012chiral}, the large quantum metric associated with the first-Landau-level-like character of the band, non-uniform quantum geometry~\cite{shavit2025quantum}, and spin anisotropy---work together to generate substantial screening of the intervalley interaction~[Fig.~\ref{fig:M4_PairingSymmetry}(c)].

We find that, once screening within the random-phase approximation is taken into account, the interaction increases with momentum, naturally favoring a sign-changing superconducting order parameter~\cite{shavit2025quantum}. Solving the linearized Bardeen--Cooper--Schrieffer gap equation, we identify the leading pairing instability~[Fig.~\ref{fig:M4_PairingSymmetry}(d)] as an sign-changing extended $s$-wave state in the $A_1$ irreducible representation, consistent with the symmetry and overall structure of the gap function obtained from our exact diagonalization calculations.

\paragraph{Field Theory for Continuous Transition at $d_{c,2}$.}\noindent
Our exact diagonalization calculations reveal a direct continuous transition from the non-Abelian fractional spin Hall insulator to an $s$-wave superconductor, where topological order and symmetry, i.e., charge conservation symmetry, change simultaneously. Here, by topological order we refer to the underlying anyon topological data~\cite{wen2015theory,kitaev2006anyons,wen2017colloquium,lan2016theory,lan2017classification,barkeshli2019symmetry,cho2023classification}. In this sense, the superconducting phase corresponds to a trivial fermionic topological order~\cite{wen2017colloquium,lan2016theory,hansson2004superconductors}, whereas the fractional spin Hall insulator realizes the $\mathrm{Pf} \times \overline{\mathrm{Pf}}$ topological order. Such a transition, lying beyond the traditional Landau–Ginzburg framework of symmetry breaking, warrants further theoretical explanation. We provide a field theoretic description of this continuous transition from the non-Abelian fractional spin Hall insulator to the superconductor, building on the $\mathrm{U}(2)_{2, -4} \times \mathrm{U}(1)_8$ Chern-Simons theory of the Pfaffian state~\cite{fradkin1998chern,seiberg2016gapped,lian2018theory,hsin2020effective}.

Because the fractional spin Hall insulator in our model is adiabatically connected to the limit $d \to +\infty$, where the two half-filled valleys decouple and each supports a Pfaffian state~\cite{reddy2024non}, the resulting topological order of the fractional spin Hall insulator is $\mathrm{Pf} \times \overline{\mathrm{Pf}}$, with $\overline{\mathrm{Pf}}$ denoting the time-reversal conjugate of the Pfaffian order. Its field theoretic description is 
\begin{align}
&\mathcal{L}_{\mathrm{Pf}\times\overline{\mathrm{Pf}}}=\mathcal{L}_\mathrm{Pf}+\mathcal{L}_{\overline{\mathrm{Pf}}}\nonumber\\
&=-\frac{2}{4\pi}\Trace\left(a\diff{a}+\frac{2}{3}a^3\right)+\frac{3}{4\pi} \Trace{a} \diff \Trace{a} +\frac{1}{2\pi}A\diff \Trace{a} \nonumber\\
&\quad+\frac{2}{4\pi}\Trace\left(b\diff{b}+\frac{2}{3}b^3\right)-\frac{3}{4\pi} \Trace{b} \diff \Trace{b} -\frac{1}{2\pi}A\diff \Trace{b} ,\nonumber 
\end{align}
where $a, b$ are the dynamical $\mathrm{U}(2)$ gauge fields and $A$ is the background electromagnetic gauge field. 

The superconducting phase emerges from the condensation of the bifundamental field $\Phi$, which is coupled to the gauge fields through $\Diff \Phi = \partial \Phi - i a \Phi - i \Phi b$. Notably, we identify the field $\Phi$ with the anyon $\sigma_{\frac{1}{4}}\otimes \sigma_{\frac{1}{4}}^{\mathcal T}$~\cite{supp}, where $\sigma_{\frac{1}{4}}$ denotes the non-Abelian charge-$e/4$ anyon in the Pfaffian theory~\cite{bruillard2017fermionic,shi2025doping_} and $\sigma_{\frac{1}{4}}^{\mathcal T}$ represents its time-reversal partner. 
This anyon $\Phi$ is self-bosonic, yet exhibits nontrivial braiding statistics with other anyons in the theory, e.g., $\psi_0\otimes \sigma_{\frac{1}{4}}^\mathcal{T}$, and possesses multiple fusion channels, i.e., it is non-Abelian~\cite{supp}. 
To model the condensation transition, we introduce the potential
\begin{align*}
V(\Phi^\dagger \Phi) = m^2 \Trace{\Phi^\dagger \Phi} + \lambda_1 \Trace{\Phi^\dagger \Phi \Phi^\dagger \Phi} + \frac{\lambda_2}{2} (\Trace{\Phi^\dagger \Phi})^2,
\end{align*}
with the real coefficients $m^2$, $\lambda_1$, and $\lambda_2$. When $m^2 < 0$, the potential $V(\Phi^\dagger \Phi)$ is minimized at $\langle \Phi \rangle = s I$ with $s > 0$. This condensation enforces the constraint $\alpha \equiv a = -b$ at low energies, leading to the effective theory
\begin{align} \label{eq:superconductor_lagrangian}
\mathcal{L}_{\mathrm{Pf}\times\overline{\mathrm{Pf}}}\to\frac{2}{2\pi}A\diff \Trace{\alpha},
\end{align}
which is precisely the effective field theory of a superconductor~\cite{hansson2004superconductors}. Several technical field-theoretic details of the condensed phase, including the fate of the $\mathrm{U}(2)$ gauge fields, the compactification of $\mathrm{Tr}\,\alpha$, and the absence of gapless Goldstone modes associated with $\Phi$, can be found in \cite{supp}.

We conclude with a few remarks. First, transitions of this type, known as Chern--Simons--Landau--Ginzburg transitions, have been extensively studied in the literature and are widely believed to be continuous~\cite{wen1993transitions,kou2009mutual,barkeshli2010anyon,barkeshli2014continuous,gazit2018confinement,zeng2020continuous,wu2023continuous,schuler2023emergent,hwang2024anyon,han2025anyon,zhou2025chern,huang2025third,wang2026emergent,ji2026self}. Our exact diagonalization results are consistent with these observations. Second, the above field-theoretic analysis indicates that the transition is driven by the condensation of the charge-$e/2$ self-bosonic non-Abelian anyon $\Phi$, identified as the composite of $\sigma_{\frac{1}{4}}$ in $\mathrm{Pf}$ and its time-reversal partner $\sigma_{\frac{1}{4}}^{\mathcal T}$ in $\overline{\mathrm{Pf}}$. This interpretation is further reinforced by the category-theoretic framework of anyon superconductivity recently developed in \cite{seo2026unified}, whose application to the present system is presented in detail in \cite{supp}. Finally, the above construction generalizes straightforwardly to the case of $\mathrm{aPf} \times \overline{\mathrm{aPf}}$, with $\mathrm{aPf}$ denoting the anti-Pfaffian order~\cite{supp}.

\section{Discussion}\noindent
In this manuscript, we theoretically uncover the remarkable emergence of superconductivity from strongly correlated topological flat bands, realized as an intermediate phase between the ferromagnetic Chern insulator and the non-Abelian fractional spin Hall insulator. This superconducting state is identified through multiple hallmarks, including negative binding energy, distinctive spectral features of the pair density matrix, and finite phase stiffness. Direct evaluation of the order parameter and the dominant wavefunction of the pair density matrix reveals its pairing symmetry to be time-reversal symmetric, extended nodal $s$-wave. We further elucidate two complementary pairing mechanisms: a Kohn--Luttinger instability arising from non-uniform quantum geometry when approached from the normal metallic side~\cite{guerci2025fractionalization,xu2025chiral,shavit2025quantum}, and the condensation of charge-$e/2$ anyons when approached from the fractional spin Hall insulator. In particular, the latter mechanism establishes our system as a realistic platform for realizing anyon superconductivity~\cite{shi2025doping,shi2025doping_,pichler2025microscopic,wang2025chiral,zhang2025holon,nosov2025anyon,shi2025anyon,huang2025apparent,zhang2025charge4e,kuhlenkamp2025robust,shi2026charge4e,seo2026unified,laughlin1988superconducting,laughlin1988relationship}.

The experimental observation of both ferromagnetism~\cite{li2025universal} and the fractional spin Hall insulator~\cite{kang2024evidence,kang2025time} in twisted bilayer MoTe${}_2$ strongly suggests that tuning between these regimes could open a pathway to realizing the intermediate superconducting phase uncovered in our analysis. More specifically, the effective screening length $d$ could be tuned through gating or substrate engineering, thereby promoting superconductivity. This follows from the two following observations. First, the length scale $d$ arises from band mixing and is roughly set by the inverse of the effective band gap between the first and second moir\'{e} bands~\cite{abouelkomsan2025non}. Second, in twisted bilayer MoTe${}_2$, the Coulomb energy is comparable to the bare band gap, so at filling $\nu_h = 3$ the effective band gap is strongly renormalized by interactions. As a result, by controlling interactions—and thereby the effective band gap—the parameter $d$ can be effectively controlled. Indeed, a rough estimate of $d$ for experimentally available devices~\cite{zhang2024polarization} places it near the transition points~\cite{supp}, suggesting that superconductivity may be within experimental reach. Additionally, larger-scale numerical studies---such as quasi-one-dimensional density-matrix renormalization group calculations---would be valuable for further mitigating the finite-size effects inherent in our exact diagonalization results and for performing systematic scaling analyses near the anyon-condensation transition~\cite{supp}, thereby yielding deeper insight into the physics of this model.

\section{Methods}
\paragraph{Hartree-Fock Band Calculations.}\noindent
{Because of computational constraints, our exact diagonalization studies are limited to a single band per valley. To account for corrections from emptying the first moir\'{e} bands and to focus on the half-filled second moir\'{e} bands ($\nu_h = 3$), we performed self-consistent Hartree–Fock calculations at $\nu_h = 2$, including the four highest moir\'{e} bands. The corresponding renormalized bands and their quantum geometries are provided in the Supplementary Materials~\cite{supp}.}

\paragraph{Exact Diagonalization.}\noindent
We make use of spin $\mathrm{U}_s(1)$, moir\'{e} translation to divide the many-body Fock space into subspaces of fixed hole number, magnetization, and momentum. When applicable, we also apply $C_2^y$, valley-reduced inversion~\cite{jia2024moire,ahn2024non}, and time-reversal symmetries. Within each subspace, the many-body Hamiltonian is represented with a sparse matrix, and diagonalized using Krylov methods.

We note that we estimate the superconducting coherence length as $\ell_\text{SC}\approx 1.83\,a_M$~\cite{supp}, which is less than the side lengths of the finite lattices used in our calculations. As such, we expect superconducting pairing to be visible, despite finite-size effects~\cite{supp}.

\vspace{10pt}\noindent
\textit{- Note Added}: While finalizing this manuscript, we became aware of a paper with some overlapping results~\cite{das2025fractional}. In this paper, the authors tuned the intervalley interaction in a different form from Eq.~\eqref{eq:interaction}, and also found a superconductor between the ferromagnetic Chern insulator and non-Abelian fractional spin Hall insulator at half-filling.

\section{Acknowledgments}\noindent 
We thank Fiona Burnell, Debanjan Chowdhury, Hee-Cheol Kim, Yong Baek Kim, Yves Kwan, Inti Sodemann, Ashvin Vishwanath, and Ya-Hui Zhang for helpful discussions. This work is financially supported by Samsung Science and Technology Foundation under Project Number SSTF-BA2401-03, the NRF of Korea (Grants No. RS-2024-00410027, RS-2023-NR119931, RS-2024-00444725, RS-2023-00256050, IRS-2025-25453111, RS-2025-08542968) funded by the Korean Government (MSIT), the Air Force Office of Scientific Research under Award No. FA23862514026, and Institute of Basic Science under project code IBS-R014-D1. This work was performed in part at Aspen Center for Physics, which is supported by National Science Foundation grant PHY-2210452. The work from DGIST was supported by the National Research Foundation of Korea (NRF) (Grant No. RS-2025-00557717, RS-2023-00269616)  and the Nano and Material Technology Development Program through the National Research Foundation of Korea (NRF) funded by Ministry of Science and ICT (No. RS-2024-00444725). We also acknowledge the partner group program of the Max Planck Society. Part of this work was supported by Global Partnership Program of Leading Universities in Quantum Science and Technology (RS-2025-02317602)

\section{Competing interests}\noindent 
The authors declare no competing interest. 

\section{Author Contributions}\noindent 
G.~Y.~C. conceived and supervised the project. C.~A. performed the Hartree–Fock analysis, exact diagonalization calculations on moir\'{e} bands, and numerical characterization of the many-body ground-states. D.~S. carried out the effective field-theory analysis and the analysis of anyon data related to anyon condensation. G.~L. performed exact diagonalization calculations on generalized first Landau levels. Y.~K. provided experimental inputs that informed the theoretical proposal. C.~A., D.~S., and G.~Y.~C. co-wrote the manuscript. All authors discussed the results and contributed to the final version of the manuscript.

\bibliography{refs.bib}

@misc{supp,
    author = {},
    title = {},
    year = {2026},
    note = "See Supplemental Information at [URL].",
}

@article{laughlin1988superconducting,
    title     = {Superconducting Ground State of Noninteracting Particles Obeying Fractional Statistics},
    author    = {Laughlin, R. B.},
    journal   = {Phys. Rev. Lett.},
    volume    = {60},
    issue     = {25},
    pages     = {2677--2680},
    numpages  = {0},
    year      = {1988},
    month     = {Jun},
    publisher = {American Physical Society},
    doi       = {10.1103/PhysRevLett.60.2677}
}

@article{laughlin1988relationship,
    title     = {The Relationship Between High-Temperature Superconductivity and the Fractional Quantum Hall Effect},
    author    = {R. B. Laughlin },
    journal   = {Science},
    volume    = {242},
    number    = {4878},
    pages     = {525-533},
    year      = {1988},
    publisher = {American Association for the Advancement of Science},
    doi       = {10.1126/science.242.4878.525}
}

@article{shi2025doping,
    title     = {Doping a Fractional Quantum Anomalous Hall Insulator},
    author    = {Shi, Zhengyan Darius and Senthil, T.},
    journal   = {Phys. Rev. X},
    volume    = {15},
    issue     = {3},
    pages     = {031069},
    numpages  = {31},
    year      = {2025},
    month     = {Sep},
    publisher = {American Physical Society},
    doi       = {10.1103/kcm5-hx56}
}

@article{shi2025doping_,
    title     = {Doping lattice non-Abelian quantum Hall states},
    author    = {Zhengyan Darius Shi and Carolyn Zhang and Senthil Todadri},
    journal   = {SciPost Phys.},
    volume    = {19},
    pages     = {150},
    year      = {2025},
    publisher = {SciPost},
    doi       = {10.21468/SciPostPhys.19.6.150},
}

@article{pichler2025microscopic,
    title     = {Microscopic mechanism of anyon superconductivity emerging from fractional Chern insulators},
    author    = {Pichler, Fabian and Kuhlenkamp, Clemens and Knap, Michael and Vishwanath, Ashvin},
    journal   = {Newton},
    year      = {2025},
    month     = {Dec},
    publisher = {Elsevier},
    doi       = {10.1016/j.newton.2025.100340}
}

@misc{wang2025chiral,
    title         = {Chiral superconductivity near a fractional Chern insulator}, 
    author        = {Taige Wang and Michael P. Zaletel},
    year          = {2025},
    eprint        = {2507.07921},
    archivePrefix = {arXiv},
    primaryClass  = {cond-mat.str-el},
    url           = {https://arxiv.org/abs/2507.07921}, 
}

@misc{zhang2025holon,
    title         = {Holon metal, charge-density-wave and chiral superconductor from doping fractional Chern insulator and $\mathrm{SU}(3)_1$ chiral spin liquid}, 
    author        = {Ya-Hui Zhang},
    year          = {2025},
    eprint        = {2506.00110},
    archivePrefix = {arXiv},
    primaryClass  = {cond-mat.str-el},
    url           = {https://arxiv.org/abs/2506.00110}, 
}

@misc{nosov2025anyon,
    title         = {Anyon superconductivity and plateau transitions in doped fractional quantum anomalous Hall insulators}, 
    author        = {Pavel A. Nosov and Zhaoyu Han and Eslam Khalaf},
    year          = {2025},
    eprint        = {2506.02108},
    archivePrefix = {arXiv},
    primaryClass  = {cond-mat.str-el},
    url           = {https://arxiv.org/abs/2506.02108}, 
}

@misc{shi2025anyon,
    title         = {Anyon delocalization transitions out of a disordered FQAH insulator}, 
    author        = {Zhengyan Darius Shi and T. Senthil},
    year          = {2025},
    eprint        = {2506.02128},
    archivePrefix = {arXiv},
    primaryClass  = {cond-mat.str-el},
    url           = {https://arxiv.org/abs/2506.02128}, 
}

@misc{huang2025apparent,
    title     = {Apparent inconsistency between Streda formula and Hall conductivity in reentrant integer quantum anomalous Hall effect in twisted {Mo}{Te}${}_2$},
    author    = {Huang, Yi and Musser, Seth and Zhu, Jihang and Chou, Yang-Zhi and Das Sarma, Sankar},
    journal   = {Phys. Rev. B},
    volume    = {112},
    issue     = {19},
    pages     = {195136},
    numpages  = {11},
    year      = {2025},
    month     = {Nov},
    publisher = {American Physical Society},
    doi       = {10.1103/ml2q-39pp},
}

@misc{zhang2025charge4e,
    title         = {Charge-4$e$ Anyon Superconductor from Doping $\text{SU}(4)_1$ chiral spin liquid}, 
    author        = {Lu Zhang and Ya-Hui Zhang and Xue-Yang Song},
    year          = {2025},
    eprint        = {2508.12370},
    archivePrefix = {arXiv},
    primaryClass  = {cond-mat.str-el},
    url           = {https://arxiv.org/abs/2508.12370}, 
}

@misc{kuhlenkamp2025robust,
    title         = {Robust superconductivity upon doping chiral spin liquid and Chern insulators in a Hubbard-Hofstadter model}, 
    author        = {Clemens Kuhlenkamp and Stefan Divic and Michael P. Zaletel and Tomohiro Soejima and Ashvin Vishwanath},
    year          = {2025},
    eprint        = {2509.02675},
    archivePrefix = {arXiv},
    primaryClass  = {cond-mat.str-el},
    url           = {https://arxiv.org/abs/2509.02675}, 
}

@misc{shi2026charge4e,
    title         = {Charge-$4e$ superconductor with parafermionic vortices: A path to universal topological quantum computation}, 
    author        = {Zhengyan Darius Shi and Zhaoyu Han and Srinivas Raghu and Ashvin Vishwanath},
    year          = {2026},
    eprint        = {2602.06963},
    archivePrefix = {arXiv},
    primaryClass  = {cond-mat.str-el},
    url           = {https://arxiv.org/abs/2602.06963}, 
}

@article{kohn1965new,
    title     = {New Mechanism for Superconductivity},
    author    = {Kohn, W. and Luttinger, J. M.},
    journal   = {Phys. Rev. Lett.},
    volume    = {15},
    issue     = {12},
    pages     = {524--526},
    numpages  = {0},
    year      = {1965},
    month     = {Sep},
    publisher = {American Physical Society},
    doi       = {10.1103/PhysRevLett.15.524}
}

@article{shavit2025quantum,
    title     = {Quantum Geometric Kohn-Luttinger Superconductivity},
    author    = {Shavit, Gal and Alicea, Jason},
    journal   = {Phys. Rev. Lett.},
    volume    = {134},
    issue     = {17},
    pages     = {176001},
    numpages  = {7},
    year      = {2025},
    month     = {Apr},
    publisher = {American Physical Society},
    doi       = {10.1103/PhysRevLett.134.176001}
}

@article{jahin2026enhanced,
    title     = {Enhanced Kohn-Luttinger superconductivity in geometric bands},
    author    = {Jahin, Ammar and Lin, Shi-Zeng},
    journal   = {Phys. Rev. B},
    volume    = {113},
    issue     = {1},
    pages     = {014504},
    numpages  = {8},
    year      = {2026},
    month     = {Jan},
    publisher = {American Physical Society},
    doi       = {10.1103/gt8h-czf3}
}

@article{maiti2010renormalization,
    title     = {Renormalization group flow, competing phases, and the structure of superconducting gap in multiband models of iron-based superconductors},
    author    = {Maiti, Saurabh and Chubukov, Andrey V.},
    journal   = {Phys. Rev. B},
    volume    = {82},
    issue     = {21},
    pages     = {214515},
    year      = {2010},
    month     = {Dec},
    publisher = {American Physical Society},
    doi       = {10.1103/PhysRevB.82.214515}
}

@article{gonzalez2008kohn,
    title     = {Kohn-Luttinger superconductivity in graphene},
    author    = {Gonz\'alez, J.},
    journal   = {Phys. Rev. B},
    volume    = {78},
    issue     = {20},
    pages     = {205431},
    numpages  = {6},
    year      = {2008},
    month     = {Nov},
    publisher = {American Physical Society},
    doi       = {10.1103/PhysRevB.78.205431}
}

@article{nandkishore2012chiral,
    title     = {Chiral superconductivity from repulsive interactions in doped graphene},
    author    = {Nandkishore, Rahul and Levitov, L. S. and Chubukov, A. V.},
    journal   = {Nature Physics},
    volume    = {8},
    issue     = {2},
    pages     = {158--163},
    year      = {2012},
    month     = {Feb},
    publisher = {Nature Publishing Group UK London},
    doi       = {10.1038/nphys2208}
}

@article{fradkin1998chern,
    title     = {A Chern-Simons effective field theory for the Pfaffian quantum Hall state},
    author    = {Eduardo Fradkin and Chetan Nayak and Alexei Tsvelik and Frank Wilczek},
    journal   = {Nuclear Physics B},
    volume    = {516},
    number    = {3},
    pages     = {704--718},
    year      = {1998},
    publisher = {Elsevier},
    doi       = {10.1016/S0550-3213(98)00111-4}
}

@article{seiberg2016gapped,
    title     = {Gapped boundary phases of topological insulators via weak coupling},
    author    = {Seiberg, Nathan and Witten, Edward},
    journal   = {Prog. Theor. Exp. Phys.},
    volume    = {2016},
    number    = {12},
    pages     = {12C101},
    year      = {2016},
    month     = {Nov},
    publisher = {Oxford University Press},
    doi       = {10.1093/ptep/ptw083}
}

@article{lian2018theory,
    title     = {Theory of the disordered $\ensuremath{\nu}=\frac{5}{2}$ quantum thermal Hall state: Emergent symmetry and phase diagram},
    author    = {Lian, Biao and Wang, Juven},
    journal   = {Phys. Rev. B},
    volume    = {97},
    issue     = {16},
    pages     = {165124},
    numpages  = {25},
    year      = {2018},
    month     = {Apr},
    publisher = {American Physical Society},
    doi       = {10.1103/PhysRevB.97.165124}
}

@article{hsin2020effective,
    title     = {Effective field theory for fractional quantum Hall systems near $\ensuremath{\nu}=5/2$},
    author    = {Hsin, Po-Shen and Lin, Ying-Hsuan and Paquette, Natalie M. and Wang, Juven},
    journal   = {Phys. Rev. Res.},
    volume    = {2},
    issue     = {4},
    pages     = {043242},
    numpages  = {31},
    year      = {2020},
    month     = {Nov},
    publisher = {American Physical Society},
    doi       = {10.1103/PhysRevResearch.2.043242}
}

@article{hansson2004superconductors,
    title     = {Superconductors are topologically ordered},
    author    = {T. H. Hansson and Vadim Oganesyan and S. L. Sondhi},
    journal   = {Annals of Physics},
    volume    = {313},
    number    = {2},
    pages     = {497–538},
    year      = {2004},
    month     = {Oct}, 
    publisher = {Elsevier},
    doi       = {10.1016/j.aop.2004.05.006}
}

@article{wen1993transitions,
    title     = {Transitions between the quantum Hall states and insulators induced by periodic potentials},
    author    = {Wen, Xiao-Gang and Wu, Yong-Shi},
    journal   = {Phys. Rev. Lett.},
    volume    = {70},
    issue     = {10},
    pages     = {1501--1504},
    numpages  = {0},
    year      = {1993},
    month     = {Mar},
    publisher = {American Physical Society},
    doi       = {10.1103/PhysRevLett.70.1501}
}

@article{kou2009mutual,
    title     = {Mutual Chern-Simons Landau-Ginzburg theory for continuous quantum phase transition of $\mathbb{Z}_{2}$ topological order},
    author    = {Kou, Su-Peng and Yu, Jing and Wen, Xiao-Gang},
    journal   = {Phys. Rev. B},
    volume    = {80},
    issue     = {12},
    pages     = {125101},
    numpages  = {5},
    year      = {2009},
    month     = {Sep},
    publisher = {American Physical Society},
    doi       = {10.1103/PhysRevB.80.125101}
}

@article{barkeshli2010anyon,
    title     = {Anyon Condensation and Continuous Topological Phase Transitions in Non-Abelian Fractional Quantum Hall States},
    author    = {Barkeshli, Maissam and Wen, Xiao-Gang},
    journal   = {Phys. Rev. Lett.},
    volume    = {105},
    issue     = {21},
    pages     = {216804},
    numpages  = {4},
    year      = {2010},
    month     = {Nov},
    publisher = {American Physical Society},
    doi       = {10.1103/PhysRevLett.105.216804}
}

@article{barkeshli2014continuous,
    title     = {Continuous transition between fractional quantum Hall and superfluid states},
    author    = {Barkeshli, Maissam and McGreevy, John},
    journal   = {Phys. Rev. B},
    volume    = {89},
    issue     = {23},
    pages     = {235116},
    numpages  = {6},
    year      = {2014},
    month     = {Jun},
    publisher = {American Physical Society},
    doi       = {10.1103/PhysRevB.89.235116}
}

@misc{zhou2025chern,
    title             = {Chern-Simons-matter conformal field theory on fuzzy sphere: Confinement transition of Kalmeyer-Laughlin chiral spin liquid}, 
    author            = {Zheng Zhou and Chong Wang and Yin-Chen He},
    year              = {2025},
    eprint            = {2507.19580},
    archivePrefix     = {arXiv},
    primaryClass      = {cond-mat.str-el},
    url               = {https://arxiv.org/abs/2507.19580}, 
}

@misc{han2025anyon,
    title         = {Anyon superfluidity of excitons in quantum Hall bilayers}, 
    author        = {Zhaoyu Han and Taige Wang and Zhihuan Dong and Michael P. Zaletel and Ashvin Vishwanath},
    year          = {2025},
    eprint        = {2508.14894},
    archivePrefix = {arXiv},
    primaryClass  = {cond-mat.str-el},
    url           = {https://arxiv.org/abs/2508.14894}, 
}

@misc{huang2025third,
    title         = {Third-order quantum phase transitions of bosonic non-Abelian fractional quantum Hall states}, 
    author        = {Kai-Wen Huang and Xiang-Jian Hou and Ying-Hai Wu},
    year          = {2025},
    eprint        = {2509.17113},
    archivePrefix = {arXiv},
    primaryClass  = {cond-mat.str-el},
    url           = {https://arxiv.org/abs/2509.17113}, 
}

@misc{ji2026self,
    title         = {Self-dual Higgs transitions: Toric code and beyond}, 
    author        = {Wenjie Ji and Ryan A. Lanzetta and Zheng Zhou and Chong Wang},
    year          = {2026},
    eprint        = {2601.20945},
    archivePrefix = {arXiv},
    primaryClass  = {cond-mat.str-el},
    url           = {https://arxiv.org/abs/2601.20945}, 
}

@article{wu2023continuous,
    title     = {Continuous Phase Transitions between Fractional Quantum Hall States and Symmetry-Protected Topological States},
    author    = {Wu, Ying-Hai and Tu, Hong-Hao and Cheng, Meng},
    journal   = {Phys. Rev. Lett.},
    volume    = {131},
    issue     = {25},
    pages     = {256502},
    numpages  = {7},
    year      = {2023},
    month     = {Dec},
    publisher = {American Physical Society},
    doi       = {10.1103/PhysRevLett.131.256502}
}

@article{gazit2018confinement,
    title = {Confinement transition of $\mathbb{Z}_2$ gauge theories coupled to massless fermions: Emergent quantum chromodynamics and $\mathrm{SO}$(5) symmetry},
    author    = {Snir Gazit  and Fakher F. Assaad  and Subir Sachdev  and Ashvin Vishwanath  and Chong Wang},
    journal   = {Proceedings of the National Academy of Sciences},
    volume    = {115},
    number    = {30},
    pages     = {E6987-E6995},
    year      = {2018},
    month     = {Jul},
    publisher = {National Academy of Sciences},
    doi       = {10.1073/pnas.1806338115},
}

@article{zeng2020continuous,
    title     = {Continuous phase transition between bosonic integer quantum Hall liquid and a trivial insulator: Evidence for deconfined quantum criticality},
    author    = {Zeng, Tian-Sheng and Sheng, D. N. and Zhu, W.},
    journal   = {Phys. Rev. B},
    volume    = {101},
    issue     = {3},
    pages     = {035138},
    numpages  = {6},
    year      = {2020},
    month     = {Jan},
    publisher = {American Physical Society},
    doi       = {10.1103/PhysRevB.101.035138}
}

@article{schuler2023emergent,
    title     = {Emergent XY* transition driven by symmetry fractionalization and anyon condensation},
    author    = {Michael Schuler and Louis-Paul Henry and Yuan-Ming Lu and Andreas M. Läuchli},
    journal   = {SciPost Phys.},
    volume    = {14},
    pages     = {001},
    year      = {2023},
    publisher = {SciPost},
    doi       = {10.21468/SciPostPhys.14.1.001}
}

@article{hwang2024anyon,
    title     = {Anyon condensation and confinement transition in a Kitaev spin liquid bilayer},
    author    = {Hwang, Kyusung},
    journal   = {Phys. Rev. B},
    volume    = {109},
    issue     = {13},
    pages     = {134412},
    numpages  = {18},
    year      = {2024},
    month     = {Apr},
    publisher = {American Physical Society},
    doi       = {10.1103/PhysRevB.109.134412}
}

@misc{wang2026emergent,
    title         = {Emergent QED${}_3$ at the bosonic Laughlin state to superfluid transition}, 
    author        = {Taige Wang and Xue-Yang Song and Michael P. Zaletel and T. Senthil},
    year          = {2026},
    eprint        = {2507.07611},
    archivePrefix = {arXiv},
    primaryClass  = {cond-mat.str-el},
    url           = {https://arxiv.org/abs/2507.07611}, 
}

@misc{seo2026unified,
    title         = {A Unified Categorical Description of Quantum Hall Hierarchy and Anyon Superconductivity}, 
    author        = {Donghae Seo and Taegon Lee and Gil Young Cho},
    year          = {2026},
    eprint        = {2602.03848},
    archivePrefix = {arXiv},
    primaryClass  = {cond-mat.str-el},
    url           = {https://arxiv.org/abs/2602.03848}
}

@article{wen2015theory,
    title     = {A theory of 2+1D bosonic topological orders},
    author    = {Wen, Xiao-Gang},
    journal   = {National Science Review},
    volume    = {3},
    number    = {1},
    pages     = {68–106},
    year      = {2015},
    month     = {Nov}, 
    publisher = {Oxford University Press},
    doi       = {10.1093/nsr/nwv077}
}

@article{kitaev2006anyons,
    title     = {Anyons in an exactly solved model and beyond},
    author    = {Kitaev, Alexei},
    journal   = {Annals of Physics},
    volume    = {321},
    number    = {1},
    pages     = {2–111},
    year      = {2006},
    month     = {Jan},
    publisher = {Elsevier},
    doi       = {10.1016/j.aop.2005.10.005},
}

@article{wen2017colloquium,
    title     = {Colloquium: Zoo of quantum-topological phases of matter},
    author    = {Wen, Xiao-Gang},
    journal   = {Rev. Mod. Phys.},
    volume    = {89},
    issue     = {4},
    pages     = {041004},
    numpages  = {17},
    year      = {2017},
    month     = {Dec},
    publisher = {American Physical Society},
    doi       = {10.1103/RevModPhys.89.041004}
}

@article{lan2016theory,
    title     = {Theory of (2+1)-dimensional fermionic topological orders and fermionic/bosonic topological orders with symmetries},
    author    = {Lan, Tian and Kong, Liang and Wen, Xiao-Gang},
    journal   = {Phys. Rev. B},
    volume    = {94},
    issue     = {15},
    pages     = {155113},
    numpages  = {27},
    year      = {2016},
    month     = {Oct},
    publisher = {American Physical Society},
    doi       = {10.1103/PhysRevB.94.155113}
}

@article{bruillard2017fermionic,
    title     = {Fermionic modular categories and the 16-fold way},
    author    = {Bruillard, Paul and Galindo, César and Hagge, Tobias and Ng, Siu-Hung and Plavnik, Julia Yael and Rowell, Eric C. and Wang, Zhenghan},
    journal   = {Journal of Mathematical Physics},
    volume    = {58},
    number    = {4},
    pages     = {041704},
    year      = {2017},
    month     = {Apr},
    publisher = {AIP Publishing},
    doi       = {10.1063/1.4982048},
}

@article{lan2017classification,
    title     = {Classification of (2+1)-dimensional topological order and symmetry-protected topological order for bosonic and fermionic systems with on-site symmetries},
    author    = {Lan, Tian and Kong, Liang and Wen, Xiao-Gang},
    journal   = {Phys. Rev. B},
    volume    = {95},
    issue     = {23},
    pages     = {235140},
    numpages  = {37},
    year      = {2017},
    month     = {Jun},
    publisher = {American Physical Society},
    doi       = {10.1103/PhysRevB.95.235140}
}

@article{cho2023classification,
    title     = {Classification of fermionic topological orders from congruence representations},
    author    = {Cho, Gil Young and Kim, Hee-Cheol and Seo, Donghae and You, Minyoung},
    journal   = {Phys. Rev. B},
    volume    = {108},
    issue     = {11},
    pages     = {115103},
    numpages  = {17},
    year      = {2023},
    month     = {Sep},
    publisher = {American Physical Society},
    doi       = {10.1103/PhysRevB.108.115103}
}

@article{barkeshli2019symmetry,
    title     = {Symmetry fractionalization, defects, and gauging of topological phases},
    author    = {Barkeshli, Maissam and Bonderson, Parsa and Cheng, Meng and Wang, Zhenghan},
    journal   = {Phys. Rev. B},
    volume    = {100},
    issue     = {11},
    pages     = {115147},
    numpages  = {99},
    year      = {2019},
    month     = {Sep},
    publisher = {American Physical Society},
    doi       = {10.1103/PhysRevB.100.115147}
}

@article{wang2021exact,
    title     = {Exact Landau Level Description of Geometry and Interaction in a Flatband},
    author    = {Wang, Jie and Cano, Jennifer and Millis, Andrew J. and Liu, Zhao and Yang, Bo},
    journal   = {Phys. Rev. Lett.},
    volume    = {127},
    issue     = {24},
    pages     = {246403},
    numpages  = {6},
    year      = {2021},
    month     = {Dec},
    publisher = {American Physical Society},
    doi       = {10.1103/PhysRevLett.127.246403}
}

@article{fujimoto2025higher,
    title     = {Higher Vortexability: Zero-Field Realization of Higher Landau Levels},
    author    = {Fujimoto, Manato and Parker, Daniel E. and Dong, Junkai and Khalaf, Eslam and Vishwanath, Ashvin and Ledwith, Patrick},
    journal   = {Phys. Rev. Lett.},
    volume    = {134},
    issue     = {10},
    pages     = {106502},
    numpages  = {8},
    year      = {2025},
    month     = {Mar},
    publisher = {American Physical Society},
    doi       = {10.1103/PhysRevLett.134.106502}
}

@article{liu2025theory,
    title     = {Theory of Generalized Landau Levels and Its Implications for Non-Abelian States},
    author    = {Liu, Zhao and Mera, Bruno and Fujimoto, Manato and Ozawa, Tomoki and Wang, Jie},
    journal   = {Phys. Rev. X},
    volume    = {15},
    issue     = {3},
    pages     = {031019},
    numpages  = {39},
    year      = {2025},
    month     = {Jul},
    publisher = {American Physical Society},
    doi       = {10.1103/1zg9-qbd6}
}

@article{iskin2023extracting,
    title     = {Extracting quantum-geometric effects from Ginzburg-Landau theory in a multiband Hubbard model},
    author    = {Iskin, M.},
    journal   = {Phys. Rev. B},
    volume    = {107},
    issue     = {22},
    pages     = {224505},
    year      = {2023},
    month     = {Jun},
    publisher = {American Physical Society},
    doi       = {10.1103/PhysRevB.107.224505}
}

@article{chen2024ginzburg,
    title     = {Ginzburg-Landau Theory of Flat-Band Superconductors with Quantum Metric},
    author    = {Chen, Shuai A. and Law, K. T.},
    journal   = {Phys. Rev. Lett.},
    volume    = {132},
    issue     = {2},
    pages     = {026002},
    numpages  = {7},
    year      = {2024},
    month     = {Jan},
    publisher = {American Physical Society},
    doi       = {10.1103/PhysRevLett.132.026002}
}

@article{hu2025anomalous,
    title     = {Anomalous coherence length in superconductors with quantum metric},
    author    = {Hu, Jin-Xin and Chen, Shuai A. and Law, K. T.},
    journal   = {Communications Physics},
    volume    = {8},
    issue     = {1},
    pages     = {20},
    year      = {2025},
    month     = {Jan},
    publisher = {Nature Publishing Group UK London},
    doi       = {10.1038/s42005-024-01930-0}
}

@article{iskin2025pair,
    title     = {Pair size and quantum geometry in a multiband Hubbard model},
    author    = {Iskin, M.},
    journal   = {Phys. Rev. B},
    volume    = {111},
    issue     = {1},
    pages     = {014502},
    numpages  = {7},
    year      = {2025},
    month     = {Jan},
    publisher = {American Physical Society},
    doi       = {10.1103/PhysRevB.111.014502}
}

@article{thumin2025correlation,
    title     = {Correlation functions and characteristic lengthscales in flat band superconductors},
    author    = {Maxime Thumin and Georges Bouzerar},
    journal   = {SciPost Phys.},
    volume    = {18},
    pages     = {025},
    year      = {2025},
    month     = {Jan},
    publisher = {SciPost},
    doi       = {10.21468/SciPostPhys.18.1.025}
}

@article{wu2019topological,
    title     = {Topological Insulators in Twisted Transition Metal Dichalcogenide Homobilayers},
    author    = {Wu, Fengcheng and Lovorn, Timothy and Tutuc, Emanuel and Martin, Ivar and MacDonald, A. H.},
    journal   = {Phys. Rev. Lett.},
    volume    = {122},
    issue     = {8},
    pages     = {086402},
    numpages  = {5},
    year      = {2019},
    month     = {Feb},
    publisher = {American Physical Society},
    doi       = {10.1103/PhysRevLett.122.086402}
}

@article{jia2024moire,
    title     = {Moir{\'e} fractional Chern insulators. I. First-principles calculations and continuum models of twisted bilayer {Mo}{Te}${}_2$},
    author    = {Jia, Yujin and Yu, Jiabin and Liu, Jiaxuan and Herzog-Arbeitman, Jonah and Qi, Ziyue and Pi, Hanqi and Regnault, Nicolas and Weng, Hongming and Bernevig, B. Andrei and Wu, Quansheng},
    journal   = {Phys. Rev. B},
    volume    = {109},
    issue     = {20},
    pages     = {205121},
    numpages  = {29},
    year      = {2024},
    month     = {May},
    publisher = {American Physical Society},
    doi       = {10.1103/PhysRevB.109.205121}
}

@article{zhang2024polarization,
    title     = {Polarization-driven band topology evolution in twisted {Mo}{Te}${}_2$ and {W}{Se}${}_2$},
    author    = {Zhang, Xiao-Wei and Wang, Chong and Liu, Xiaoyu and Fan, Yueyao and Cao, Ting and Xiao, Di},
    journal   = {Nature Communications},
    volume    = {15},
    number    = {1},
    pages     = {4223},
    year      = {2024},
    month     = {May},
    publisher = {Nature Publishing Group UK London},
    doi       = {10.1038/s41467-024-48511-x}
}

@article{xiao2012coupled,
    title     = {Coupled Spin and Valley Physics in Monolayers of {Mo}{S}${}_2$ and Other Group-VI Dichalcogenides},
    author    = {Xiao, Di and Liu, Gui-Bin and Feng, Wanxiang and Xu, Xiaodong and Yao, Wang},
    journal   = {Phys. Rev. Lett.},
    volume    = {108},
    issue     = {19},
    pages     = {196802},
    numpages  = {5},
    year      = {2012},
    month     = {May},
    publisher = {American Physical Society},
    doi       = {10.1103/PhysRevLett.108.196802}
}

@article{shi2024adiabatic,
    title     = {Adiabatic approximation and Aharonov-Casher bands in twisted homobilayer transition metal dichalcogenides},
    author    = {Shi, Jingtian and Morales-Dur\'an, Nicol\'as and Khalaf, Eslam and MacDonald, A. H.},
    journal   = {Phys. Rev. B},
    volume    = {110},
    issue     = {3},
    pages     = {035130},
    numpages  = {17},
    year      = {2024},
    month     = {Jul},
    publisher = {American Physical Society},
    doi       = {10.1103/PhysRevB.110.035130}
}

@article{guerci2025fractionalization,
    title     = {From Fractionalization to Chiral Topological Superconductivity in a Flat Chern Band},
    author    = {Guerci, Daniele and Abouelkomsan, Ahmed and Fu, Liang},
    journal   = {Phys. Rev. Lett.},
    volume    = {135},
    issue     = {18},
    pages     = {186601},
    numpages  = {8},
    year      = {2025},
    month     = {Oct},
    publisher = {American Physical Society},
    doi       = {10.1103/zm39-dstj}
}

@article{xu2025chiral,
    title     = {Chiral Superconductivity from Spin Polarized Chern Band in Twisted {Mo}{Te}${}_2$},
    author    = {Xu, Cheng and Zou, Nianlong and Peshcherenko, Nikolai and Jahin, Ammar and Li, Tingxin and Lin, Shi-Zeng and Zhang, Yang},
    journal   = {Phys. Rev. Lett.},
    volume    = {135},
    issue     = {26},
    pages     = {266005},
    year      = {2025},
    month     = {Dec},
    publisher = {American Physical Society},
    doi       = {10.1103/h22z-4hsj}
}

@article{ahn2024non,
    title     = {Non-Abelian fractional quantum anomalous Hall states and first Landau level physics of the second moir{\'e} band of twisted bilayer {Mo}{Te}${}_2$},
    author    = {Ahn, Cheong-Eung and Lee, Wonjun and Yananose, Kunihiro and Kim, Youngwook and Cho, Gil Young},
    journal   = {Phys. Rev. B},
    volume    = {110},
    issue     = {16},
    pages     = {L161109},
    year      = {2024},
    month     = {Oct},
    publisher = {American Physical Society},
    doi       = {10.1103/PhysRevB.110.L161109}
}

@article{reddy2024non,
    title     = {Non-Abelian Fractionalization in Topological Minibands},
    author    = {Reddy, Aidan P. and Paul, Nisarga and Abouelkomsan, Ahmed and Fu, Liang},
    journal   = {Phys. Rev. Lett.},
    volume    = {133},
    issue     = {16},
    pages     = {166503},
    numpages  = {8},
    year      = {2024},
    month     = {Oct},
    publisher = {American Physical Society},
    doi       = {10.1103/PhysRevLett.133.166503}
}

@article{chen2025robust,
    title     = {Robust non-Abelian even-denominator fractional Chern insulator in twisted bilayer {Mo}{Te}${}_2$},
    author    = {Chen, Feng and Luo, Wei-Wei and Zhu, Wei and Sheng, DN},
    journal   = {Nature Communications},
    volume    = {16},
    number    = {1},
    pages     = {2115},
    year      = {2025},
    publisher = {Nature Publishing Group UK London},
    doi       = {10.1038/s41467-025-57326-3}
}

@article{xu2025multiple,
    title     = {Multiple Chern Bands in Twisted {Mo}{Te}${}_2$ and Possible Non-Abelian States},
    author    = {Xu, Cheng and Mao, Ning and Zeng, Tiansheng and Zhang, Yang},
    journal   = {Phys. Rev. Lett.},
    volume    = {134},
    issue     = {6},
    pages     = {066601},
    numpages  = {7},
    year      = {2025},
    month     = {Feb},
    publisher = {American Physical Society},
    doi       = {10.1103/PhysRevLett.134.066601}
}

@article{wang2025higher,
    title     = {Higher Landau-Level Analogs and Signatures of Non-Abelian States in Twisted Bilayer {Mo}{Te}${}_2$},
    author    = {Wang, Chong and Zhang, Xiao-Wei and Liu, Xiaoyu and Wang, Jie and Cao, Ting and Xiao, Di},
    journal   = {Phys. Rev. Lett.},
    volume    = {134},
    issue     = {7},
    pages     = {076503},
    numpages  = {7},
    year      = {2025},
    month     = {Feb},
    publisher = {American Physical Society},
    doi       = {10.1103/PhysRevLett.134.076503}
}

@article{liu2025non,
    title     = {Non-Abelian Fractional Chern Insulators and Competing States in Flat Moir{\'e} Bands},
    author    = {Liu, Hui and Liu, Zhao and Bergholtz, Emil J.},
    journal   = {Phys. Rev. Lett.},
    volume    = {135},
    issue     = {10},
    pages     = {106604},
    numpages  = {8},
    year      = {2025},
    month     = {Sep},
    publisher = {American Physical Society},
    doi       = {10.1103/43nq-ntqm}
}

@article{kwan2026regarding,
    title = {Regarding the existence of abelian fractional topological insulators in twisted {Mo}{Te}${}_2$ and related systems},
    author    = {Kwan, Yves H. and Wagner, Glenn and Yu, Jiabin and Dagnino, Andrea Kouta and Jiang, Yi and Xu, Xiaodong and Bernevig, B. Andrei and Neupert, Titus and Regnault, Nicolas},
    journal   = {Communications Physics},
    volume    = {9},
    issue     = {1},
    year      = {2026},
    month     = {Jan},
    publisher = {Nature Publishing Group UK London},
    doi       = {10.1038/s42005-025-02483-6}
}

@article{yang1962concept,
    title     = {Concept of Off-Diagonal Long-Range Order and the Quantum Phases of Liquid He and of Superconductors},
    author    = {Yang, C. N.},
    journal   = {Rev. Mod. Phys.},
    volume    = {34},
    issue     = {4},
    pages     = {694--704},
    numpages  = {0},
    year      = {1962},
    month     = {Oct},
    publisher = {American Physical Society},
    doi       = {10.1103/RevModPhys.34.694}
}

@article{scalapino1993insulator,
    title     = {Insulator, metal, or superconductor: The criteria},
    author    = {Scalapino, Douglas J. and White, Steven R. and Zhang, Shoucheng},
    journal   = {Phys. Rev. B},
    volume    = {47},
    issue     = {13},
    pages     = {7995--8007},
    numpages  = {0},
    year      = {1993},
    month     = {Apr},
    publisher = {American Physical Society},
    doi       = {10.1103/PhysRevB.47.7995}
}

@article{resta2018drude,
    title     = {Drude weight and superconducting weight},
    author    = {Resta, Raffaele},
    journal   = {Journal of Physics: Condensed Matter},
    volume    = {30},
    number    = {41},
    pages     = {414001},
    year      = {2018},
    month     = {Sep},
    publisher = {IOP Publishing},
    doi       = {10.1088/1361-648X/aade19}
}

@article{kang2024evidence,
    title     = {Evidence of the fractional quantum spin Hall effect in moir{\'e} {Mo}{Te}${}_2$},
    author    = {Kang, Kaifei and Shen, Bowen and Qiu, Yichen and Zeng, Yihang and Xia, Zhengchao and Watanabe, Kenji and Taniguchi, Takashi and Shan, Jie and Mak, Kin Fai},
    journal   = {Nature},
    volume    = {628},
    issue     = {8008},
    pages     = {522--526},
    year      = {2024},
    month     = {Apr},
    doi       = {10.1038/s41586-024-07214-5},
    publisher = {Nature Publishing Group UK London}
}

@misc{kang2025time,
    title         = {Time-reversal symmetry breaking fractional quantum spin Hall insulator in moir{\'e} {Mo}{Te}${}_2$}, 
    author        = {Kaifei Kang and Yichen Qiu and Bowen Shen and Kihong Lee and Zhengchao Xia and Yihang Zeng and Kenji Watanabe and Takashi Taniguchi and Jie Shan and Kin Fai Mak},
    year          = {2025},
    eprint        = {2501.02525},
    archivePrefix = {arXiv},
    primaryClass  = {cond-mat.mes-hall},
    url           = {https://arxiv.org/abs/2501.02525}, 
}

@article{li2025universal,
    title     = {Universal Magnetic Phases in Twisted Bilayer {Mo}{Te}${}_2$},
    author    = {Li, Weijie and Redekop, Evgeny and Wang Beach, Christiano and Zhang, Canxun and Zhang, Xiaowei and Liu, Xiaoyu and Holtzmann, Will and Hu, Chaowei and Anderson, Eric and Park, Heonjoon and Taniguchi, Takashi and Watanabe, Kenji and Chu, Jiun-haw and Fu, Liang and Cao, Ting and Xiao, Di and Young, Andrea F. and Xu, Xiaodong},
    journal   = {Nano Letters},
    volume    = {25},
    number    = {52},
    pages     = {18044--18050},
    year      = {2025},
    publisher = {ACS Publications},
    doi       = {10.1021/acs.nanolett.5c04751}
}

@misc{xu2025signatures,
    title         = {Signatures of unconventional superconductivity near reentrant and fractional quantum anomalous Hall insulators}, 
    author        = {Fan Xu and Zheng Sun and Jiayi Li and Ce Zheng and Cheng Xu and Jingjing Gao and Tongtong Jia and Kenji Watanabe and Takashi Taniguchi and Bingbing Tong and Li Lu and Jinfeng Jia and Zhiwen Shi and Shengwei Jiang and Yuanbo Zhang and Yang Zhang and Shiming Lei and Xiaoxue Liu and Tingxin Li},
    year          = {2025},
    eprint        = {2504.06972},
    archivePrefix = {arXiv},
    primaryClass  = {cond-mat.mes-hall},
    url           = {https://arxiv.org/abs/2504.06972}, 
}

@article{cao2018unconventional,
    title     = {Unconventional superconductivity in magic-angle graphene superlattices},
    author    = {Cao, Yuan and Fatemi, Valla and Fang, Shiang and Watanabe, Kenji and Taniguchi, Takashi and Kaxiras, Efthimios and Jarillo-Herrero, Pablo},
    journal   = {Nature},
    volume    = {556},
    number    = {7699},
    pages     = {43--50},
    year      = {2018},
    month     = {Apr},
    publisher = {Nature Publishing Group},
    doi       = {10.1038/nature26160}
}

@article{chen2019signatures,
    title     = {Signatures of tunable superconductivity in a trilayer graphene moir{\'e} superlattice},
    author    = {Chen, Guorui and Sharpe, Aaron L. and Gallagher, Patrick and Rosen, Ilan T. and Fox, Eli J. and Jiang, Lili and Lyu, Bosai and Li, Hongyuan and Watanabe, Kenji and Taniguchi, Takashi and Jung, Jeil and Shi, Zhiwen and Goldhaber-Gordon, David and Zhang, Yuanbo and Wang, Feng},
    journal   = {Nature},
    volume    = {572},
    number    = {7768},
    pages     = {215--219},
    year      = {2019},
    month     = {Oct},
    publisher = {Nature Publishing Group UK London},
    doi       = {10.1038/s41586-019-1393-y}
}

@article{oh2021evidence,
    title     = {Evidence for unconventional superconductivity in twisted bilayer graphene},
    author    = {Oh, Myungchul and Nuckolls, Kevin P. and Wong, Dillon and Lee, Ryan L. and Liu, Xiaomeng and Watanabe, Kenji and Taniguchi, Takashi and Yazdani, Ali},
    journal   = {Nature},
    volume    = {600},
    issue     = {7888},
    pages     = {240--245},
    year      = {2021},
    month     = {Dec},
    publisher = {Nature Publishing Group UK London},
    doi       = {10.1038/s41586-021-04121-x}
}

@article{tian2023evidence,
    title     = {Evidence for Dirac flat band superconductivity enabled by quantum geometry},
    author    = {Tian, Haidong and Gao, Xueshi and Zhang, Yuxin and Che, Shi and Xu, Tianyi and Cheung, Patrick and Watanabe, Kenji and Taniguchi, Takashi and Randeria, Mohit and Zhang, Fan and Lau, Chun Ning and Bockrath, Marc W.},
    journal   = {Nature},
    volume    = {614},
    issue     = {7948},
    pages     = {440--444},
    year      = {2023},
    month     = {Feb},
    publisher = {Nature Publishing Group UK London},
    doi       = {10.1038/s41586-022-05576-2}
}

@article{park2021tunable,
    title     = {Tunable strongly coupled superconductivity in magic-angle twisted trilayer graphene},
    author    = {Park, Jeong Min and Cao, Yuan and Watanabe, Kenji and Taniguchi, Takashi and Jarillo-Herrero, Pablo},
    journal   = {Nature},
    volume    = {590},
    issue     = {7845},
    pages     = {249--255},
    year      = {2021},
    month     = {Feb},
    publisher = {Nature Publishing Group UK London},
    doi       = {10.1038/s41586-021-03192-0}
}

@article{kim2022evidence,
    title     = {Evidence for unconventional superconductivity in twisted trilayer graphene},
    author    = {Kim, Hyunjin and Choi, Youngjoon and Lewandowski, Cyprian and Thomson, Alex and Zhang, Yiran and Polski, Robert and Watanabe, Kenji and Taniguchi, Takashi and Alicea, Jason and Nadj-Perge, Stevan},
    journal   = {Nature},
    volume    = {606},
    issue     = {7914},
    pages     = {494--500},
    year      = {2022},
    month     = {Jun},
    publisher = {Nature Publishing Group UK London},
    doi       = {10.1038/s41586-022-04715-z}
}

@article{xu2023observation,
    title     = {Observation of Integer and Fractional Quantum Anomalous Hall Effects in Twisted Bilayer {Mo}{Te}${}_2$},
    author    = {Xu, Fan and Sun, Zheng and Jia, Tongtong and Liu, Chang and Xu, Cheng and Li, Chushan and Gu, Yu and Watanabe, Kenji and Taniguchi, Takashi and Tong, Bingbing and Jia, Jinfeng and Shi, Zhiwen and Jiang, Shengwei and Zhang, Yang and Liu, Xiaoxue and Li, Tingxin},
    journal   = {Phys. Rev. X},
    volume    = {13},
    issue     = {3},
    pages     = {031037},
    numpages  = {12},
    year      = {2023},
    month     = {Sep},
    publisher = {American Physical Society},
    doi       = {10.1103/PhysRevX.13.031037}
}

@article{cai2023signatures,
    title     = {Signatures of fractional quantum anomalous Hall states in twisted {Mo}{Te}${}_2$},
    author    = {Cai, Jiaqi and Anderson, Eric and Wang, Chong and Zhang, Xiaowei and Liu, Xiaoyu and Holtzmann, William and Zhang, Yinong and Fan, Fengren and Taniguchi, Takashi and Watanabe, Kenji and Ran, Ying and Cao, Ting and Fu, Liang and Xiao, Di and Yao, Wang and Xu, Xiaodong},
    journal   = {Nature},
    volume    = {622},
    number    = {7981},
    pages     = {63--68},
    year      = {2023},
    month     = {Oct},
    publisher = {Nature Publishing Group UK London},
    doi       = {10.1038/s41586-023-06289-w}
}

@article{zeng2023thermodynamic,
    title     = {Thermodynamic evidence of fractional Chern insulator in moir{\'e} {Mo}{Te}${}_2$},
    author    = {Zeng, Yihang and Xia, Zhengchao and Kang, Kaifei and Zhu, Jiacheng and Kn{\"u}ppel, Patrick and Vaswani, Chirag and Watanabe, Kenji and Taniguchi, Takashi and Mak, Kin Fai and Shan, Jie},
    journal   = {Nature},
    volume    = {622},
    number    = {7981},
    pages     = {69--73},
    year      = {2023},
    month     = {Oct},
    publisher = {Nature Publishing Group UK London},
    doi       = {10.1038/s41586-023-06452-3}
}

@article{park2023observation,
    title     = {Observation of fractionally quantized anomalous Hall effect},
    author    = {Park, Heonjoon and Cai, Jiaqi and Anderson, Eric and Zhang, Yinong and Zhu, Jiayi and Liu, Xiaoyu and Wang, Chong and Holtzmann, William and Hu, Chaowei and Liu, Zhaoyu and Taniguchi, Takashi and Watanabe, Kenji and Chu, Jiun-Haw and Cao, Ting and Fu, Liang and Yao, Wang and Chang, Cui-Zu and Cobden, David and Xiao, Di and Xu, Xiaodong},
    journal   = {Nature},
    volume    = {622},
    number    = {7981},
    pages     = {74--79},
    year      = {2023},
    month     = {Oct},
    publisher = {Nature Publishing Group UK London},
    doi       = {10.1038/s41586-023-06536-0}
}

@article{ji2024local,
    title     = {Local probe of bulk and edge states in a fractional Chern insulator},
    author    = {Ji, Zhurun and Park, Heonjoon and Barber, Mark E. and Hu, Chaowei and Watanabe, Kenji and Taniguchi, Takashi and Chu, Jiun-Haw and Xu, Xiaodong and Shen, Zhi-Xun},
    journal   = {Nature},
    volume    = {635},
    number    = {8039},
    pages     = {578--583},
    year      = {2024},
    month     = {Nov},
    publisher = {Nature Publishing Group UK London},
    doi       = {10.1038/s41586-024-08092-7}
}

@article{redekop2024direct,
    title     = {Direct magnetic imaging of fractional Chern insulators in twisted {Mo}{Te}${}_2$},
    author    = {Redekop, Evgeny and Zhang, Canxun and Park, Heonjoon and Cai, Jiaqi and Anderson, Eric and Sheekey, Owen and Arp, Trevor and Babikyan, Grigory and Salters, Samuel and Watanabe, Kenji and Taniguchi, Takashi and Huber, Martin E. and Xu, Xiaodong and Young, Andrea F.},
    journal   = {Nature},
    volume    = {635},
    number    = {8039},
    pages     = {584--589},
    year      = {2024},
    month     = {Nov},
    publisher = {Nature Publishing Group UK London},
    doi       = {10.1038/s41586-024-08153-x}
}

@misc{park2025observation,
    title         = {Observation of High-Temperature Dissipationless Fractional Chern Insulator}, 
    author        = {Heonjoon Park and Weijie Li and Chaowei Hu and Christiano Beach and Miguel Gonçalves and Juan Felipe Mendez-Valderrama and Jonah Herzog-Arbeitman and Takashi Taniguchi and Kenji Watanabe and David Cobden and Liang Fu and B. Andrei Bernevig and Nicolas Regnault and Jiun-Haw Chu and Di Xiao and Xiaodong Xu},
    year          = {2025},
    eprint        = {2503.10989},
    archivePrefix = {arXiv},
    primaryClass  = {cond-mat.mes-hall},
    url           = {https://arxiv.org/abs/2503.10989}, 
}

@article{lu2024fractional,
    title     = {Fractional quantum anomalous Hall effect in multilayer graphene},
    author    = {Lu, Zhengguang and Han, Tonghang and Yao, Yuxuan and Reddy, Aidan P. and Yang, Jixiang and Seo, Junseok and Watanabe, Kenji and Taniguchi, Takashi and Fu, Liang and Ju, Long},
    journal   = {Nature},
    volume    = {626},
    number    = {8000},
    pages     = {759--764},
    year      = {2024},
    month     = {Feb},
    publisher = {Nature Publishing Group UK London},
    doi       = {10.1038/s41586-023-07010-7}
}

@article{lu2025extended,
    title     = {Extended quantum anomalous Hall states in graphene/hBN moir{\'e} superlattices},
    author    = {Lu, Zhengguang and Han, Tonghang and Yao, Yuxuan and Hadjri, Zach and Yang, Jixiang and Seo, Junseok and Shi, Lihan and Ye, Shenyong and Watanabe, Kenji and Taniguchi, Takashi and Ju, Long},
    journal   = {Nature},
    volume    = {637},
    number    = {8048},
    pages     = {1090--1095},
    year      = {2025},
    month     = {Jan},
    publisher = {Nature Publishing Group UK London},
    doi       = {10.1038/s41586-024-08470-1}
}

@article{xie2025tunable,
    title     = {Tunable fractional Chern insulators in rhombohedral graphene superlattices},
    author    = {Xie, Jian and Huo, Zihao and Lu, Xin and Feng, Zuo and Zhang, Zaizhe and Wang, Wenxuan and Yang, Qiu and Watanabe, Kenji and Taniguchi, Takashi and Liu, Kaihui and Song, Zhida and Xie, X. C. and Liu, Jianpeng and Lu, Xiaobo},
    journal   = {Nature Materials},
    volume    = {24},
    number    = {7},
    pages     = {1--7},
    year      = {2025},
    month     = {Jul},
    publisher = {Nature Publishing Group UK London},
    doi       = {10.1038/s41563-025-02225-7}
}

@article{aronson2025displacement,
    title     = {Displacement Field-Controlled Fractional Chern Insulators and Charge Density Waves in a Graphene/hBN Moir{\'e} Superlattice},
    author    = {Aronson, Samuel H. and Han, Tonghang and Lu, Zhengguang and Yao, Yuxuan and Butler, Jackson P. and Watanabe, Kenji and Taniguchi, Takashi and Ju, Long and Ashoori, Raymond C.},
    journal   = {Phys. Rev. X},
    volume    = {15},
    issue     = {3},
    pages     = {031026},
    numpages  = {9},
    year      = {2025},
    month     = {Jul},
    publisher = {American Physical Society},
    doi       = {10.1103/75gl-jzl6}
}

@article{cao2018correlated,
    title     = {Correlated insulator behaviour at half-filling in magic-angle graphene superlattices},
    author    = {Cao, Yuan and Fatemi, Valla and Demir, Ahmet and Fang, Shiang and Tomarken, Spencer L. and Luo, Jason Y. and Sanchez-Yamagishi, Javier D. and Watanabe, Kenji and Taniguchi, Takashi and Kaxiras, Efthimios and Ashoori, Ray C. and Jarillo-Herrero, Pablo},
    journal   = {Nature},
    volume    = {556},
    number    = {7699},
    pages     = {80--84},
    year      = {2018},
    month     = {Apr},
    publisher = {Nature Publishing Group UK London},
    doi       = {10.1038/nature26154}
}

@article{regan2020mott,
    title={Mott and generalized Wigner crystal states in {W}{Se}{${}_2$}/{W}{S}{${}_2$} moir{\'e} superlattices},
    author    = {Regan, Emma C. and Wang, Danqing and Jin, Chenhao and Bakti Utama, M. Iqbal and Gao, Beini and Wei, Xin and Zhao, Sihan and Zhao, Wenyu and Zhang, Zuocheng and Yumigeta, Kentaro and Blei, Mark and Carlström, Johan D. and Watanabe, Kenji and Taniguchi, Takashi and Tongay, Sefaattin and Crommie, Michael and Zettl, Alex and Wang, Feng},
    journal   = {Nature},
    volume    = {579},
    number    = {7799},
    pages     = {359--363},
    year      = {2020},
    month     = {Mar},
    publisher = {Nature Publishing Group UK London},
    doi       = {10.1038/s41586-020-2092-4}
}

@article{cao2020strange,
    title     = {Strange Metal in Magic-Angle Graphene with near Planckian Dissipation},
    author    = {Cao, Yuan and Chowdhury, Debanjan and Rodan-Legrain, Daniel and Rubies-Bigorda, Oriol and Watanabe, Kenji and Taniguchi, Takashi and Senthil, T. and Jarillo-Herrero, Pablo},
    journal   = {Phys. Rev. Lett.},
    volume    = {124},
    issue     = {7},
    pages     = {076801},
    numpages  = {7},
    year      = {2020},
    month     = {Feb},
    publisher = {American Physical Society},
    doi       = {10.1103/PhysRevLett.124.076801}
}

@article{nayak2008non,
    title     = {Non-Abelian anyons and topological quantum computation},
    author    = {Nayak, Chetan and Simon, Steven H. and Stern, Ady and Freedman, Michael and Das Sarma, Sankar},
    journal   = {Rev. Mod. Phys.},
    volume    = {80},
    issue     = {3},
    pages     = {1083--1159},
    numpages  = {0},
    year      = {2008},
    month     = {Sep},
    publisher = {American Physical Society},
    doi       = {10.1103/RevModPhys.80.1083}
}

@article{abouelkomsan2025non,
    title     = {Non-Abelian spin Hall insulator},
    author    = {Abouelkomsan, Ahmed and Fu, Liang},
    journal   = {Phys. Rev. Res.},
    volume    = {7},
    issue     = {2},
    pages     = {023083},
    numpages  = {11},
    year      = {2025},
    month     = {Apr},
    publisher = {American Physical Society},
    doi       = {10.1103/PhysRevResearch.7.023083}
}

@article{shi2008phase,
    title     = {Phase diagram for bilayer quantum Hall effect at total filling ${\ensuremath{\nu}}_{T}=5$},
    author    = {Shi, Chuntai and Jolad, Shivakumar and Regnault, Nicolas and Jain, Jainendra K.},
    journal   = {Phys. Rev. B},
    volume    = {77},
    issue     = {15},
    pages     = {155127},
    numpages  = {9},
    year      = {2008},
    month     = {Apr},
    publisher = {American Physical Society},
    doi       = {10.1103/PhysRevB.77.155127}
}

@article{zhu2019exciton,
    title     = {Exciton condensation in quantum Hall bilayers at total filling ${\ensuremath{\nu}}_{T}=5$},
    author    = {Zhu, Zheng and Jian, Shao-Kai and Sheng, D. N.},
    journal   = {Phys. Rev. B},
    volume    = {99},
    issue     = {20},
    pages     = {201108},
    numpages  = {6},
    year      = {2019},
    month     = {May},
    publisher = {American Physical Society},
    doi       = {10.1103/PhysRevB.99.201108}
}

@article{oshikawa2007topological,
    title     = {Topological degeneracy of non-Abelian states for dummies},
    author    = {Oshikawa, Masaki and Kim, Yong Baek and Shtengel, Kirill and Nayak, Chetan and Tewari, Sumanta},
    journal   = {Annals of Physics},
    volume    = {322},
    number    = {6},
    pages     = {1477--1498},
    year      = {2007},
    publisher = {Elsevier},
    doi       = {10.1016/j.aop.2006.08.001}
}

@article{emery1995importance,
    title     = {Importance of phase fluctuations in superconductors with small superfluid density},
    author    = {Emery, V. J. and Kivelson, S. A.},
    journal   = {Nature},
    volume    = {374},
    number    = {6521},
    pages     = {434--437},
    year      = {1995},
    publisher = {Nature Publishing Group UK London},
    doi       = {10.1038/374434a0}
}

@article{chen2024superconductivity,
    title     = {When superconductivity crosses over: From BCS to BEC},
    author    = {Chen, Qijin and Wang, Zhiqiang and Boyack, Rufus and Yang, Shuolong and Levin, K.},
    journal   = {Rev. Mod. Phys.},
    volume    = {96},
    issue     = {2},
    pages     = {025002},
    numpages  = {54},
    year      = {2024},
    month     = {May},
    publisher = {American Physical Society},
    doi       = {10.1103/RevModPhys.96.025002}
}

@misc{das2025fractional,
    title         = {Fractional topological insulators at odd-integer filling: Phase diagram of two-valley quantum Hall model}, 
    author        = {Sahana Das and Glenn Wagner and Titus Neupert},
    year          = {2025},
    eprint        = {2509.16335},
    archivePrefix = {arXiv},
    primaryClass  = {cond-mat.str-el},
    url           = {https://arxiv.org/abs/2509.16335}, 
}

@article{polyakov1977quark,
    author = {A. M. Polyakov},
    title = {Quark confinement and topology of gauge theories},
    journal = {Nuclear Physics B},
    year = {1977},
    volume = {120},
    issue = {3},
    pages = {429--458},
    month = {mar},
    doi = {https://doi.org/10.1016/0550-3213(77)90086-4}
}

@article{teper1993confining,
    author = {M. Teper},
    title = {{The confining string and its tension for SU(2) gauge fields in 2 + 1 dimensions}},
    journal = {Physical Letters B},
    year = {1993},
    volume = {311},
    issue = {1--4},
    pages = {223--229},
    month = {jul},
    doi = {https://doi.org/10.1016/0370-2693(93)90559-Z}
}

\end{document}

\title{Supplementary Information for ``Superconductivity Proximate to Non-Abelian Fractional Spin Hall Insulator in Twisted Bilayer MoTe$_2$'' }

\author{Cheong-Eung Ahn}
\altaffiliation{These authors contributed equally to this work.}
\affiliation{Department of Physics, Pohang University of Science and Technology, Pohang, 37673, Republic of Korea}

\author{Donghae Seo}
\altaffiliation{These authors contributed equally to this work.}
\affiliation{Department of Physics, Pohang University of Science and Technology, Pohang, 37673, Republic of Korea}

\author{Gyeoul Lee}
\affiliation{Department of Physics, Ulsan National Institute of Science and Technology, Ulsan 44919, Republic of Korea}

\author{Youngwook Kim}
\affiliation{Department of Physics and Chemistry, Daegu Gyeongbuk Institute of Science and Technology (DGIST), Daegu 42988, Republic of Korea}

\author{Gil Young Cho}
\thanks{gilyoungcho@kaist.ac.kr}
\affiliation{Department of Physics, Korea Advanced Institute of Science and Technology, Daejeon 34141, Republic of Korea}
\affiliation{Center for Artificial Low Dimensional Electronic Systems, Institute for Basic Science, Pohang 37673, Korea}

\maketitle

\tableofcontents

\begin{figure}[b!]
\centering
\includegraphics[width=0.99\linewidth]{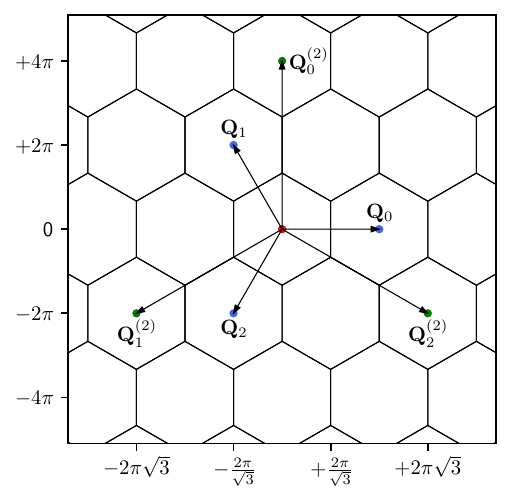}
\caption{\textbf{Moir\'{e} Brillouin Zone and Reciprocal Lattice.} The reciprocal lattice is generated by $\mathbf{Q}_1$ and $\mathbf{Q}_2$. We also label $\mathbf{Q}_0=-\mathbf{Q}_1-\mathbf{Q}_2$ and $\mathbf{Q}_j^{(2)}=\mathbf{Q}_{j+1}-\mathbf{Q}_{j+2}$ ($j\in\mathbb{Z}/3\mathbb{Z}$).}
\label{fig:S_InfiniteLattice}
\end{figure}

\section{Additional Details on the Model}
\paragraph{Lattice Geometry} We denote the moir\'{e} Brillouin zone as $\mathrm{BZ}$ and the moir\'{e} reciprocal lattice as $\hat{\Lambda}$. See [Fig.~\ref{fig:S_InfiniteLattice}] for orientation and our choice of reciprocal lattice vectors, $\mathbf{Q}_1$ and $\mathbf{Q}_2$.

\paragraph{Continuum Model of Twisted Bilayer MoTe${}_2$} Let us discuss the parameters used in our paper. We used the normalization convention of setting the moir\'{e} lattice constant as unity, $a_M=a_0\left(2\sin\frac{\theta}{2}\right)^{-1}=1$, where $a_0=3.52\:\text{\AA}$ is the monolayer lattice constant and $\theta=2.1^\circ$ is the twist angle. For the continuum model, we use the effective mass $m^*/m_e=0.62$, and consider the moir\'{e} potential up to valley-reduced-inversion symmetric second-harmonics~\cite{jia2024moire,ahn2024non},
\begin{align*}
v_\ell\left(\mathbf{r}\right)&=2v_1\sum_{j=0}^2\cos\left(\mathbf{Q}_j\cdot\mathbf{r}+\ell\psi\right)\\
&\qquad+2v_2\sum_{j=0}^2\cos\left(\mathbf{Q}_j^{(2)}\cdot\mathbf{r}\right),\\
\gamma\left(\mathbf{r}\right)&=\gamma_1\left(1+e^{+i\mathbf{Q}_2\cdot\mathbf{r}}+e^{-i\mathbf{Q}_1\cdot\mathbf{r}}\right)\\
&\qquad+\gamma_2\left(e^{-i\mathbf{Q}_0\cdot\mathbf{r}}+e^{-i\mathbf{Q}_0^{(2)}\cdot\mathbf{r}}+e^{+i\mathbf{Q}_0\cdot\mathbf{r}}\right),
\end{align*}
for some real parameters $v_1,v_2,\psi,\gamma_1,\gamma_2$. The moir\'{e} reciprocal lattice vectors $\mathbf{Q}_j$ and $\mathbf{Q}_j^{(2)}$ are as defined in [Fig.~\ref{fig:S_InfiniteLattice}].

A summary of continuum model parameter values for the second moir\'{e} band (2MB) at $\theta\sim 2^\circ$ are listed in [Table~\ref{tab:S_ContinuumModelParameters}]. For our paper, we use parameters, which have been optimized to better stabilize the non-Abelian fractional spin Hall insulator at $\theta=2.1^\circ$. We note that while all parameter sets do stabilize the fractional quantum anomalous Hall insulator in a single valley at larger system sizes, and thus are also likely to stabilize the fractional spin Hall insulator in the thermodynamic limit, it is obscured by finite size effects for the system sizes we consider, requiring the more optimized parameters.

\begin{table}[t!]
\centering
\begin{tabular}{ccccc|c}
$v_1$ & $\psi$ & $v_2$ & $\gamma_1$ & $\gamma_2$ & Ref.\\
\hline
20.51 & $-61.49^\circ$ & $-9.08$ & $-7.01$ & 11.08 & \cite{ahn2024non,zhang2024polarization}\\
17.5 & $-58.49^\circ$ & $-10.5$ & $-6.5$ & 11.75 & \cite{chen2025robust}\\
2.4 & $-90.0^\circ$ & 1.0 & $-5.8$ & 2.8 & \cite{xu2025multiple}\\
27.45 & $-58.69^\circ$ & $-6.50$ & $-12.06$ & $18.64$ & this paper\\
27.71 & $-59.08^\circ$ & $-6.44$ & $-12.02$ & $18.55$ & [Fig.~\ref{fig:S_ManyBodySpectrum_even}(b)] only
\end{tabular}
\caption{\textbf{Continuum model parameters for the 2MB at $\theta\sim 2^\circ$.} The parameters $v_1,v_2,\gamma_1,\gamma_2$ are in meV units.}
\label{tab:S_ContinuumModelParameters}
\end{table}

We denote the moir\'{e} bands with $\left(s,\nu,\mathbf{k}\right)$, where $s\in\{\uparrow,\downarrow\}$ is the spin, $\nu$ is the band index, and $\mathbf{k}\in\mathrm{BZ}$ is the moir\'{e} crystal momentum. That is,
\begin{equation*}
\hat{h}\ket{s,\nu,\mathbf{k}}=\epsilon^{s,\nu}_{\mathbf{k}}\ket{s,\nu,\mathbf{k}},
\end{equation*}
where $\epsilon^{s,\nu}_{\mathbf{k}}$ are the moir\'{e} band energies and $\ket{s,\nu,\mathbf{k}}$ are the corresponding single-electron states. We perform a particle-hole transformation and henceforth work with hole operators.

\paragraph{Interacting Hamiltonian} Let us rewrite the interacting Hamiltonian in the moir\'{e} band basis. The many-body Hamiltonian can be written as
\begin{align*}
&H=\sum_{s,\nu,\mathbf{k}}\epsilon^{s,\nu}_{\mathbf{k}}\psi^{s,\nu\dagger}_{\mathbf{k}}\psi^{s,\nu}_{\mathbf{k}}\\
&+\frac{1}{2A}\sum_{\substack{s_0,s_1,\\\nu_0,\nu_1,\nu_0',\nu_1',\\\mathbf{k}_0,\mathbf{k}_1,\mathbf{p}}}\mathfrak{V}^{s_0s_1,\nu_0'\nu_1'\nu_0,\nu_1}_{\mathbf{k}_0\mathbf{k}_1\mathbf{p}}\psi^{s_0,\nu_0'\dagger}_{\left\{\mathbf{k}_0+\mathbf{p}\right\}}\psi^{s_1,\nu_1'\dagger}_{\left\{\mathbf{k}_1-\mathbf{p}\right\}}\psi^{s_1,\nu_1}_{\mathbf{k}_1}\psi^{s_0,\nu_0}_{\mathbf{k}_0},
\end{align*}
where $\psi^{s,\nu\dagger}_{\mathbf{k}}$ are creation operators for the moir\'{e} band states $\ket{s,\nu,\mathbf{k}}$, and $\mathfrak{V}^{s_0s_1,\nu_0'\nu_1'\nu_0,\nu_1}_{\mathbf{k}_0\mathbf{k}_1\mathbf{p}}$ is the two-body interaction matrix,
\begin{align}
&\mathfrak{V}^{s_0s_1,\nu_0'\nu_1'\nu_0,\nu_1}_{\mathbf{k}_0\mathbf{k}_1\mathbf{p}}\label{eq:interaction-matrix}\\
&=\sum_{\mathbf{Q}\in\hat{\Lambda}}V^{s_0s_1}\left(\mathbf{p}+\mathbf{Q}\right)\mathfrak{f}^{s_0,\nu_0'\nu_0}_{\mathbf{k}_0,\mathbf{p}+\mathbf{Q}}\overline{\mathfrak{f}}^{s_1,\nu_1\nu_1'}_{\left\{\mathbf{k}_1-\mathbf{p}\right\},\mathbf{p}+\mathbf{Q}},\nonumber
\end{align}
where $V^{s_0s_1}(\mathbf{P})$ is specified in the main text, and $\mathfrak{f}^{s,\nu'\nu}_{\mathbf{k},\mathbf{P}}$ is the form factor,
\begin{equation*}
\mathfrak{f}^{s,\nu'\nu}_{\mathbf{k},\mathbf{P}}=\mel{s,\nu',\left\{\mathbf{k}+\mathbf{P}\right\}}{e^{+i\mathbf{P}\cdot\hat{\mathbf{r}}}}{s,\nu,\mathbf{k}}.
\end{equation*}
Here, $\{\bullet\}$ denotes the crystal momentum part of momentum, i.e. momentum modulo reciprocal lattice momentum, for some consistent prescription of the Brillouin zone, $\mathrm{BZ}$, embedded in $\mathbb{R}^2=\mathrm{BZ}\oplus\hat{\Lambda}$.

For our paper, we use $\epsilon_r=5$, which gives an interaction strength scale of $e^2/4\pi\epsilon a_M\approx 29.99\:\mathrm{meV}$.

\begin{figure}[b!]
\centering
\includegraphics[width=0.99\linewidth]{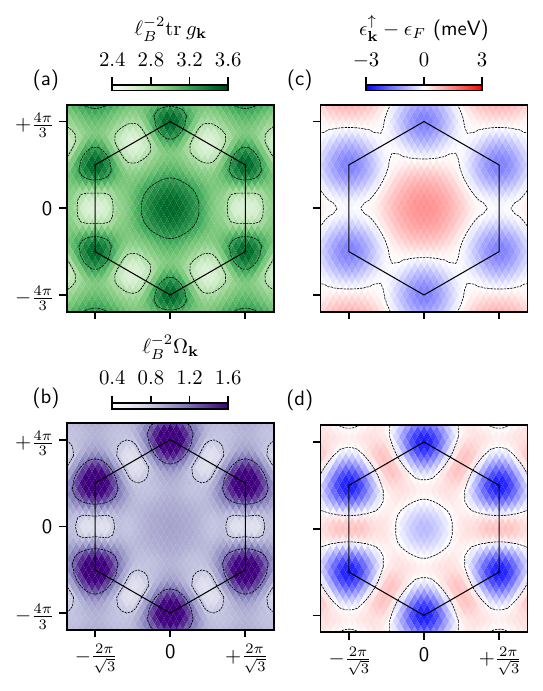}
\caption{\textbf{Hartree-Fock Renormalized Second Moir\'{e} Band at $d/a_M=0.28$.} (a) Trace of Fubini-Study metric in units of $\ell_B^{-2}$. (b) Berry curvature in units of $\ell_B^{-2}$. (c) Band energies of the second moir\'{e} band relative to the Fermi energy at half-filling, together with the Fermi surface. (d) Equivalent of (c) for the particle-hole transformed electron band.}
\label{fig:S_Bands}
\end{figure}

\paragraph{Hartree-Fock Renormalized Second Moir\'{e} Bands} In [Fig.~\ref{fig:S_Bands}], we show the Hartree-Fock renormalized second moir\'{e} bands at $\nu_h=2$ used as the background for our many-body calculations, taking $d/a_M=0.28$ as a representative example. The second moir\'{e} bands are very flat, with a bandwidth $\delta_{\text{HF},2}=2.80\:\mathrm{meV}\ll e^2/4\pi\epsilon a_M$, and fairly uniform Berry curvature and trace of Fubini-Study metric, satisfying $\frac{1}{2\pi}\int_{\mathrm{BZ}}\diff^2\mathbf{k}\:\trace{g_{\mathbf{k}}}=3.02$, close to the ideal value of $3$ for the first Landau level.

While flat relative to the interaction scale, let us still consider the $\nu_h=3$ Fermi surface. When considering the Fermi surface at $\nu_h=3$, rather than considering the second moir\'{e} bands at $\nu_h=2$, we can also equally consider the particle-hole transformed second moir\'{e} band at $\nu_h=4$. Note that as we only perform second moir\'{e} band projected calculations, here we refer to the particle-hole transformation of the isolated $\nu_h=2$ renormalized second moir\'{e} band, not the full Hartree-Fock renormalization for $\nu_h=4$. In addition to being more dispersive, the $\nu_h=4$ band splits the Fermi surface into three $\gamma$- and $\kappa$-point centered surfaces.

We note the presence of van Hove singularities at the $\mu$ points, which are close to the $\nu_h=3$ Fermi surface for both cases, and change occupancy across the particle-hole transformation. As such, we can expect the $\mu$ points to be relevant for near-Fermi surface physics.

\paragraph{Estimation of $d$} Following \cite{abouelkomsan2025non}, we obtained estimates for $d$ by comparing the largest magnitudes of the intervalley and intravalley two-particle spectra. For the purposes of estimating $d$, we approximate the twisted bilayer MoTe${}_2$ system with Landau levels and its' time-reversal conjugate. Comparing the ratios for the phenomenological interaction and band mixing considered up to second-order perturbations, we have the relation,
\begin{align*}
&\dfrac{1-\dfrac{\pi}{2^{1/2}\cdot 3^{5/4}}\kappa}{1-\dfrac{\pi}{3^{5/4}\cdot 5}\kappa}\\
&=\left(1+\frac{2}{3}\frac{d^2}{\ell_B^2}+\frac{1}{3}\frac{d^4}{\ell_B^4}\right)e^{d^2/2\ell_B^2}\mathrm{erfc}\left(\frac{d}{\sqrt{2}\ell_B}\right)\\
&\qquad-\frac{2}{3\sqrt{2\pi}}\left(\frac{d}{\ell_B}+\frac{d^3}{\ell_B^3}\right),
\end{align*}
where $\kappa=(e^2/4\pi a_M)/\Delta$ is the Coulomb interaction strength relative to the band gap, and $\ell_B$ is the effective magnetic length, $\ell_B=\sqrt{\sqrt{3}/4\pi}a_M\approx(0.37)a_M$.

When using the DFT bands at $\nu_h=0$~\cite{zhang2024polarization}, the gap between the moir\'{e} bands is only $\Delta_\text{NI}\approx 10\:\mathrm{meV}$, which corresponds to $\delta\approx 3.0$. This clearly exceeds the limits of second-order perturbation theory, with the left-hand side taking on a negative value, which cannot be attained even at $d\to\infty$. When accounting for Hartree-Fock renormalization, the effective band gap is enhanced, giving approximately $\Delta_\text{HF}=20\:\mathrm{meV}$ or $50\:\mathrm{meV}$, depending on whether we set the background at $\nu_h=2$ or $\nu_h=4$. We can expect the effective band gap at $\nu_h=3$ to be somewhere between these two values, giving us the estimate range $\kappa\in(0.6,1.5)$. Substituting into the ratio equation, we have $d/a_M\in(0.13,1.7)$, which includes our estimates for $d_{c,1}$ and $d_{c,2}$ in the possible range. We note that while Hartree-Fock renormalization tends to overestimate band gaps in renormalized bands, we can still expect $d$ to be close to the transitional values identified in our main text.

\paragraph{Estimation of Superconducting Coherence Length} Here, we estimate the superconducting coherence length at $d/a_M=0.28$.

First, we consider the conventional BCS contribution arising from the band dispersion relative to the pairing gap along the Fermi surface, $\ell_{\text{BCS}}=\hbar v_F/\Delta$, which requires estimates of both the Fermi velocity and the pairing gap. When using the hole bands, native to our calculations, the average Fermi velocity is $\hbar v_F=\left(0.67(25)\:\mathrm{meV}\right)a_M$ [Fig.~\ref{fig:S_Bands}(c)]. Due to the non-uniform quantum geometry, the second moir\'{e} band is not particle-hole symmetric and the particle-hole transformed band has a larger Fermi velocity, $\hbar v_F=\left(1.21(33)\:\mathrm{meV}\right)a_M$ [Fig.~\ref{fig:S_Bands}(d)]. We use the latter for the estimate for the superconducting coherence length. In lieu of the pairing gap, we use half of the binding energy, which for $N=16$ is found to be $|E_b| = 1.37\:\mathrm{meV}$. Thus, we have our estimate for the conventional BCS superconducting coherence length, $\ell_\text{BCS}=1.77(48)\,a_M$.

Second, as we are considering bands with non-trivial quantum geometry, there are also quantum geometric contributions to the superconducting coherence length~\cite{iskin2023extracting,chen2024ginzburg,hu2025anomalous,iskin2025pair,thumin2025correlation},  $\ell_{\text{QM}}$, which combine with the conventional BCS contribution to give the complete mean-field superconducting coherence length~\cite{hu2025anomalous},
\begin{equation*}
\ell_\text{SC}^{(\text{MF})}=\sqrt{\ell_\text{BCS}^2+\ell_\text{QM}^2}.
\end{equation*}
To obtain an approximate estimate of the contribution arising from the quantum geometry, we model the second moir\'{e} band by the first Landau level~\cite{ahn2024non}. Within this framework, the quantum geometric contribution is given by $\ell_{\mathrm{QM}} \approx \left(\det g_\text{1LL}\right)^{1/4} = \sqrt{\frac{3}{2}}\ell_B=\sqrt{\frac{3\sqrt{3}}{8\pi}}a_M\approx 0.45\,a_M$ where $g_\text{1LL}$ is the Fubini-Study metric of the first Landau level, and $\ell_B$ denotes the effective magnetic length.

Combining the two contributions~\cite{hu2025anomalous}, the resulting mean-field estimate for the superconducting coherence length is given by $\ell_{\mathrm{SC}}^{(\text{MF})}=1.83(46)\,a_M$, which is dominated by the conventional BCS contribution.

We note that as a mean-field estimate, this does not incorporate the effects of interaction, despite being the dominant energy scale. In fact, as mentioned in the main text and elaborated later in Section II.F, the superconductivity we observe is likely in the BCS-BEC crossover regime so cannot be purely described at the mean field level. While we do not carry out independent estimates incorporating interactions due to the limitations inherent to our methodology, we note that numerical studies which have studied the BCS-BEC crossover with quantum geometry effects noted a suppression of the superconducting coherence length with increasing interaction strength, which is typical of the crossover regardless of quantum geometry effects~\cite{chen2024superconductivity}, but interestingly, that the coherence length remains bounded below by $\sim\ell_\text{QM}$, even as we approach the BEC limit~\cite{iskin2025pair,thumin2025correlation}, albeit with some debate on conventions for the coherence length in the strongly-correlated limit. As such, we expect our true effective coherence length to lie somewhere along the interval $\ell_\text{SC}\in\left(\ell_\text{QM},\ell_\text{SC}^{(\text{MF})}\right)=\left(0.45,1.83(46)\right)a_M$.

\begin{figure}[b!]
\centering
\includegraphics[width=0.99\linewidth]{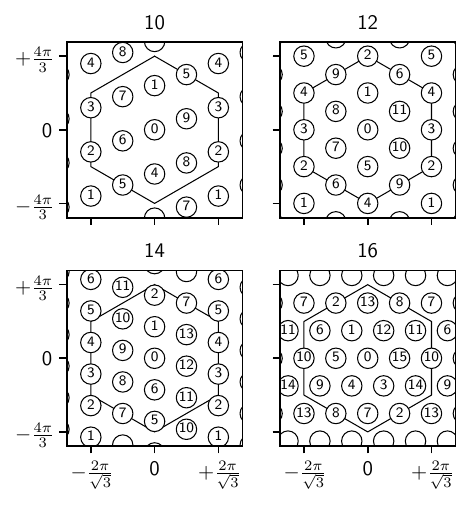}
\caption{\textbf{Brillouin Zone of Finite Size Triangular Lattices.} The hexagon in the background is the continuum Brillouin zone corresponding to an infinite triangular lattice. The numbered circles are the allowed crystal momenta of a discrete Brillouin zone corresponding to the finite size triangular lattice of sizes $N=10,12,14,16$.}
\label{fig:S_FiniteLattices}
\end{figure}

\paragraph{Finite Lattices and Finite-Size Effects} For our exact diagonalization (ED) calculations, we used finite size periodic lattices of system sizes $N=12,14,16$ [Fig.~\ref{fig:S_FiniteLattices}]. The total real space area is given by $A=\left(\sqrt{3}/2\right)a_M^2N$. Comparing the area to the superconducting coherence length,
\begin{equation*}
\frac{A}{\ell_\text{SC}^2}\gtrsim\frac{6\sqrt{3}}{\left(1.83(46)\right)^2}\gtrsim 1.98,
\end{equation*}
even at the lowest estimate. Furthermore, the shortest periodic length in each configuration is $L_\text{min}/a_M=2\sqrt{3},2\sqrt{3},4$, which nearly doubles $\ell_\text{SC}^{(\text{MF})}$. As such, we can reasonably expect pairing to develop in our ED calculations, albeit subject to finite size effects.

We make note of some additional features of our finite lattices which may contribute to the generalizability of our results to the thermodynamic limit. First, while the continuum Brillouin zone in the thermodynamic limit is $C_3^z$ symmetric, only $N=12$ and $16$ are prescribed to be $C_3^z$ symmetric. This is why we have excluded $N=14$ for expressing pairing symmetry, and we can expect this to have a slight impact on finite-size calculations.

Additionally, the points with the greatest differences in band energy are the $\gamma$ and $\kappa$ points~[Fig.~\ref{fig:S_Bands}(c)], and these, together with the $\mu$ points, are also the points with the greatest discrepancies in quantum geometry from the first Landau level~[Fig.~\ref{fig:S_Bands}(a,b)]. As such, which finite lattices manage to incorporate these points also have an impact on finite-size calculations. Notably, as we go from $N=12,14,16$, the difference between the minimum and maximum band energy for the allowed momentum points decreases. This competes with the conventional assumptions of finite size scaling where increasing $N$ should occur with fixed or increasing effective bandwidth, and may contribute towards non-monotonic trends in our behavior over $N$.

Finally, we note that another complication in finite-size scaling may be the even-odd effect of the fractional spin Hall insulator (FSHI). As a general trend in our calculations, we observe that $N=16$ shows similar behavior to $N=12$, which is distinct from $N=14$, with the latter typically exhibiting a sharper transition, e.g. sharper slopes, stronger peaks. Due to the different ground state degeneracy, the many-body interactions must manifest in the microscopic level repulsion statistics in distinct ways for even and odd cases, which may create distinct scaling behaviors.

In summary, while we estimate that our ED calculations to be sufficient to ascertain the existence of superconductivity and make an overall assessment of its' properties, larger scale numerical calculations will be highly valuable to obtain a more accurate understanding of the transition in the thermodynamic limit.

\begin{figure}[b!]
\centering
\includegraphics[width=0.99\linewidth]{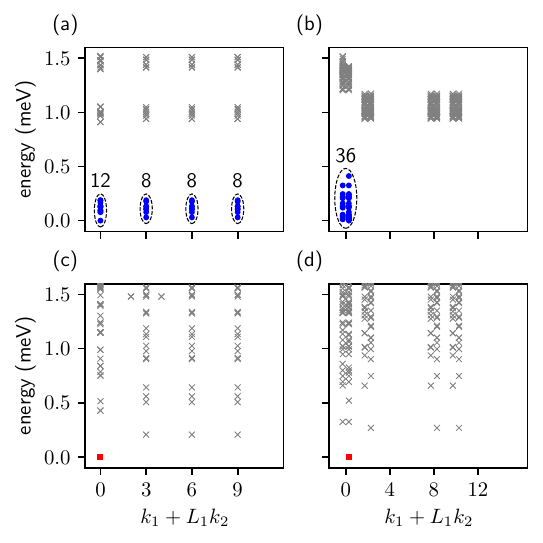}
\caption{\textbf{Many Body Spectrum for $N=12$ and $16$.} (a,b) Many-body spectra for $d/a_M=0.80$, $N=12$ and $16$, respectively. (c,d) Many-body spectrum for $d/a_M=0.28$, $N=12$ and $16$, respectively. The horizontal splits for (b,d) represent charges with respect to additional symmetries used in our computations.}
\label{fig:S_ManyBodySpectrum_even}
\end{figure}

\section{Additional Numerical Results}
\subsection{Additional Many-Body Spectra}
In the main text, we presented many-body spectra for the FSHI and superconductor for $N=14$~[Fig.~2(a,c)]. In this subsection, we present representative many-body spectra for $N=12$ and $16$. For $d/a_M=0.80$, we observe 36-fold ground state degeneracy in both $N=12$ and $16$~[Fig.~\ref{fig:S_ManyBodySpectrum_even}(a,b)], opposed to the 4-fold ground state degeneracy for $N=14$, consistent with the even-odd effect expected for the FSHI~\cite{oshikawa2007topological}. Because all 36 ground states are at $\mathbf{K}=0$ for (b), they are more strongly affected by finite-size effects, and required further optimized parameters to stabilize the FSHI [Table.~\ref{tab:S_ContinuumModelParameters}]. For $d/a_M=0.28$, we observe 1-fold ground state degeneracy in both $N=12$ and $16$~[Fig.~\ref{fig:S_ManyBodySpectrum_even}(c,d)], same as $N=14$.

\subsection{Additional Signatures of the FSHI-SC Transition}
In the main text, as evidence of the FSHI-SC transition, we focused on the FSHI ground state manifold, tracking its' ground state degeneracy and periodicity of spectral flow. In this subsection, we further quantify the evolution of many-body topology by computing the many-body spin Hall conductivity. We also provide independent and direct numerical evidence of the FSHI-SC transition by looking at the ground state energy and wavefunction fidelity susceptibilities.

\begin{figure}[b!]
\centering
\includegraphics{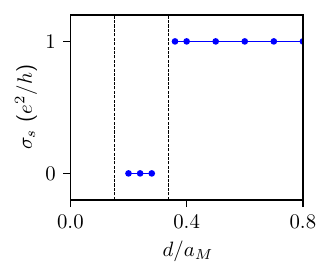}
\caption{\textbf{Many-Body Spin Hall Conductivity.} Computed for $N=14$. Dashed lines are estimated transition points from ground state energy and wavefunction fidelity susceptibilities.}
\label{fig:S_MBSpinHallConductivity}
\end{figure}

\paragraph{Many-Body Spin Hall Conductivity} We computed the many-body spin Hall conductivity from the response of the ground state wavefunction with respect to the flux insertion,
\begin{equation*}
\sigma_\text{sH}=\frac{e^2}{h}C_s,\qquad C_s=\frac{1}{2\pi\:\mathrm{GSD}}\int_{\mathbb{T}^2_{\bm{\theta}}}\diff^2\bm{\theta}\:\trace{\Omega^{(s)}_{\bm{\theta}}},
\end{equation*}
where $\mathrm{GSD}$ is the ground state degeneracy, $2\pi\theta_i=\Phi_i/\Phi_0$ is the normalized flux, $\Phi_0=h/e$ is the flux quanta, and $\Omega^{(s)}_{\bm{\theta}}$ is the many-body Berry curvature of the ground state manifold over flux, which we have computed using a $25\times 25$ grid. Note that the flux is coupled oppositely for opposite spins to give opposing current responses. If the single lowest energy state is isolated from the other states, which we judge by checking that the wavefunction fidelity for all of the links across the grid are maintained above $0.75$, we use the single lowest energy state with $\mathrm{GSD}=1$. If it requires mixing with the next highest state, we used $\mathrm{GSD}=2$. If non-trivial mixing with even higher states is required, we have excluded it from calculations. This was the case for $d/a_M=0.16$ and $0.32$, which we can associate with transition points. As noted in the main text, we observe a crossover from a many-body spin Hall conductivity of $\sigma_\text{sH}=e^2/h$ for $d>d_{c,2}$, consistent with the FSHI, to $\sigma_\text{sH}=0$ for $d<d_{c,2}$, consistent with a topologically trivial superconductor~[Fig.~\ref{fig:S_MBSpinHallConductivity}].

\begin{figure}[b!]
\centering
\includegraphics[width=0.99\linewidth]{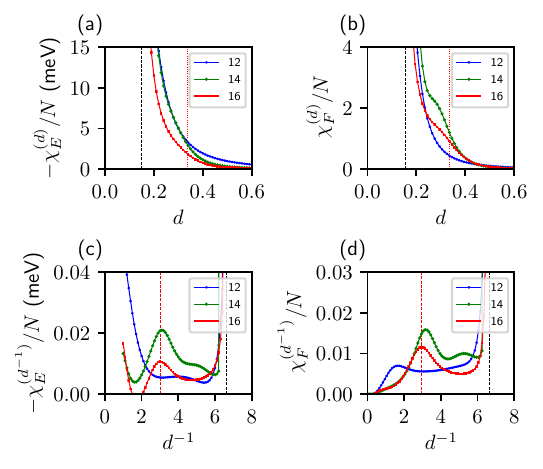}
\caption{\textbf{Ground State Energy and Wavefunction Fidelity Susceptibilities.} (a) Energy susceptibility and (b) Fidelity susceptibility of the lowest energy state in the $\mathbf{K}=0$ sector as a function of $d$. (c,d) Same as (a,b) but as a function of $d^{-1}$. Across all (a-d), we identify $d_{c,1}\approx 0.15$ (black dashed line). For (c,d), we identify peaks at $d_{c,2}\approx 0.34$ (red dashed line). This transition point is marked on (a,b) as a red dotted line, and can be associated with a non-analytic feature of the fidelity susceptibility.}
\label{fig:S_Susceptibilities}
\end{figure}

\paragraph{Ground State Energy and Wavefunction Fidelity Susceptibilities} To identify possible transitions, we wish to consider the ground state energy and wavefunction susceptibilities,
\begin{equation*}
\chi_E^{(u)}=\frac{\partial^2E_{\text{GS}}}{\partial u^2},
\end{equation*}
and
\begin{equation*}
\chi_F^{(u)}=\lim_{\delta u\rightarrow 0^+}\frac{1-\left|\braket{\text{GS};u-(\delta u/2)}{\text{GS};u+(\delta u/2)}\right|^2}{(\delta u)^2},
\end{equation*}
where $E_{\text{GS}}(u)$ is the ground state energy as a function of $u$, $\ket{\text{GS},u}$ is the ground state wavefunction as a function of $u$, and $u(d)$ is the unknown scaling field.

As we lack the system sizes to perform systematic finite size scaling, and locally determine $u(d)$, we first work with $u=d$. Note that we use finite-difference approximations with $\delta d=0.01$ to approximate the derivatives. For both the energy and fidelity susceptibilities, we can clearly identify a first-order transition at $d_{c,1}/a_M\approx 0.15$, which dominates the entire profile~[Fig.~\ref{fig:S_Susceptibilities}(a,b)]. While no clear feature presents itself for the energy susceptibility, for the fidelity susceptibility, we can see a slight kink develop for $N=14$ and $16$ at $d_{c,2}/a_M\approx 0.34$ which warrants further investigation.

Note that $d$ only enters the Hamiltonian through $e^{-d}$. As such, variations of physical observables with respect to $d$ become progressively suppressed at larger $d$. Even more, since $d_{c,2} > d_{c,1}$, and the transition at $d_{c,1}$ is significantly stronger (first order) than $d_{c,2}$, the comparatively weaker signatures near $d_{c,2}$ are more easily obscured in a direct $d$-based representation. As such, let us attempt a reparameterization of the Hamiltonian and compute the susceptibilities using $u=d^{-1}$.

Let us now look at the susceptibilites using $u=d^{-1}$. We use finite-difference approximations with $\delta d^{-1}=0.10$ to approximate the derivatives. For both the energy and fidelity susceptibilities, we can now observe a peak for $N=14$ and $16$ at $d_{c,2}/a_M\approx 0.34$~[Fig.~\ref{fig:S_Susceptibilities}(c,d)], consistent with the change in many-body spin Hall conductivity. The behavior is consistent with a continuous, likely second-order transition.

\begin{figure}[b!]
\centering
\includegraphics{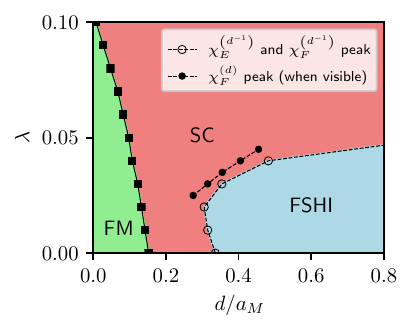}
\caption{\textbf{Extended Phase Diagram.} The FSHI-SC phase boundary is obtained using both the susceptibilities with respect to $d^{-1}$ (empty black circles) and susceptibilities with respect to $d$ (filled black circles). Only $\chi_F^{(d)}$ shows peaks for the latter, and only for the range $\lambda\in(0.025,0.045)$.}
\label{fig:S_V01-PhaseDiagram}
\end{figure}

\subsection{Extended Phase Space with Subtracted $V_0^{(1)}$}
In this subsection, we consider the extended phase space with subtracted $V_0^{(1)}$ from the intervalley interaction. As a generalization to Eq.~(2), we subtract the 0-th Haldane pseudopotential in the first Landau level contribution to the intervalley interaction,
\begin{equation*}
V^{\uparrow\downarrow}\left(\mathbf{p}\right)=\frac{e^2}{2\epsilon}\left[\frac{e^{-pd}}{p}-\lambda\ell_B f\left(\frac{d}{\ell_B}\right)\left(1-\frac{(\ell_Bp)^2}{2}\right)^{-2}\right],
\end{equation*}
where $\lambda>0$ is a dimensionless tuning parameter, and
\begin{equation*}
f(a)=\int_0^\infty\diff x\:\left(1-\frac{x^2}{2}\right)^2e^{-x^2-ax},
\end{equation*}
is a dimensionless function which monotonically decreases from $f(0)=\frac{11\sqrt{\pi}}{32}$ to $\lim_{a\rightarrow\infty}f(a)=0$. We note that the equivalent of the parameter $\lambda$ can be realized in the first moir\'{e} band by dielectric screening~\cite{kwan2026regarding}.

As this interaction does not affect the intravalley interaction while suppressing the intervalley interaction, it greatly suppresses the valley polarized IQHI relative to FSHI, while not having a significant impact on FSHI stability. It may have a small positive impact on FSHI stability at $d$ close to $d_{c,2}$, especially at finite sizes, but this would be difficult to judge without more knowledge of the wavefunction. We do however expect the 0-th Haldane pseudopotential to have a significant impact on the stability of the superconductor. As the 0th-Haldane pseudopotential is predominantly the onsite interaction, its' subtraction generates an attractive BCS $s$-wave channel for the superconducting pairing to further minimize. As our pairing gap was evaluated to be the extended $s$-wave, it can naturally hybridize with this channel to enhance the stability of the superconductor, and as we increase $\lambda$, the pairing gap will naturally evolve into a fully gapped $s$-wave.

Calculating the extended phase diagram, we observe that even a small reduction of $\lambda<0.1$ can completely suppress the ferromagnetic Chern insulator, and while it initially benefits the FSHI for $\lambda\lesssim 0.02$, it rapidly enhances the superconducting region for $\lambda\gtrsim 0.02$, quickly pushing the FSHI completely out of the experimentally relevant range of $d$~[Fig.~\ref{fig:S_V01-PhaseDiagram}].

\begin{figure}[b!]
\centering
\includegraphics[width=0.99\linewidth]{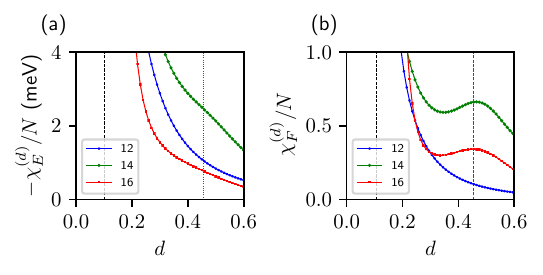}
\caption{\textbf{Ground State Energy and Wavefunction Fidelity Susceptibilities at $\lambda=0.045$.} (a) Energy susceptibility and (b) Fidelity susceptibility of the lowest energy state in the $\mathbf{K}=0$ sector as a function of $d$. We can identify a peak in the fidelity susceptibility at $d_{c,2}\approx 0.455$ (red dashed line in (b)). An associated dotted line is shown in (a) where a slight corresponding kink is visible.}
\label{fig:S_Susceptibilities_wV01-il-0045}
\end{figure}

Moreover, as $\lambda$ increases, the transition becomes more pronounced. Competing with the reduced effect of $d$ on the interaction at larger $d$ that we noted earlier, for $\lambda\in(0.025,0.045)$, we were able to directly observe peaks in the fidelity susceptibility which we attribute to $d_{c,2}$. As an example, we plot the susceptibilities for $\lambda=0.045$~[Fig.~\ref{fig:S_Susceptibilities_wV01-il-0045}]. While these estimates for $d_{c,2}$ do not exactly coincide with those obtained from the susceptibilities with respect to $d^{-1}$ due to finite size effects, their similarity is maximized at $\lambda\approx 0.03$~[Fig.~\ref{fig:S_V01-PhaseDiagram}], where the competing effects of the enhanced transition with $\lambda$ and decreased scaling of $d$ are likely maximized.

\begin{figure}[b!]
\centering
\includegraphics[width=0.99\linewidth]{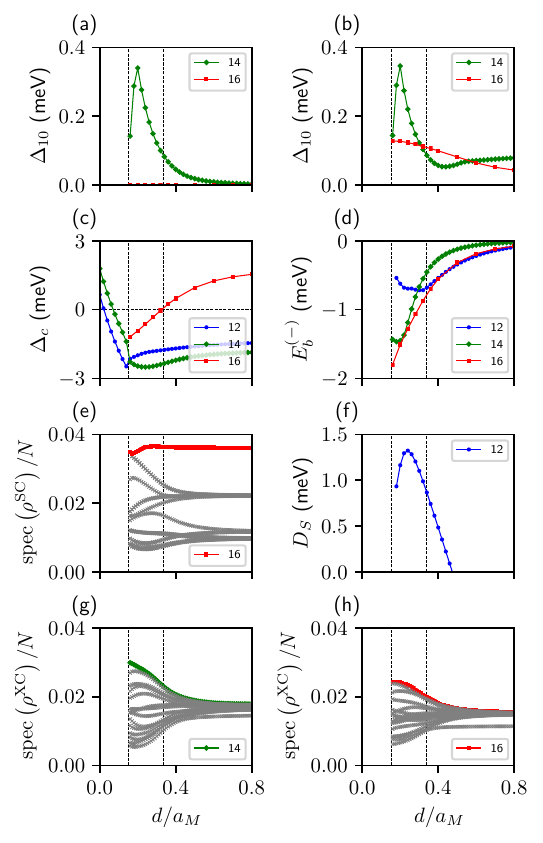}
\caption{\textbf{Additional Signatures of Superconductivity.} (a) Many-body gap $\Delta_{01}=E_1-E_0$ at $N_h=N-2$ filling. (b) Same as (a) at $N_h=N+2$ filling. (c) Charge gap. (d) Binding energy, $E_b^{(-)}$. (e) Spectrum of the pair density matrix for $N=16$. (f) Superfluid stiffness for $N=12$. (g) Spectrum of the excitonic pair density matrix for $N=14$. (h) Same as (g) for $N=16$.}
\label{fig:S_AdditionalSCSignatures}
\end{figure}

\subsection{Additional Signatures of Superconductivity}
In this subsection, we supplement the signatures of superconductivity in the intermediate regime. First, we directly address the broken $\mathrm{U}(1)$ symmetry of superconductivity by checking if superconductivity is also stabilized at neighboring fillings $N_h=N\pm 2$, and by looking at the charge gap. Next, we supplement the binding energy $E_b^{(+)}$~[Fig.~3(b)] by presenting $E_b^{(-)}$, and the pair density matrix spectra for $N=14$~[Fig.~3(c)] by presenting the same for $N=16$. Then, we supplement the superfluid stiffness for $N=14$~[Fig.~3(d)] by presenting the same for $N=12$. Finally, we exclude exciton condensation as a possible explanation for the intermediate phase by looking at the spectra of excitonic pair density matrices.

\paragraph{Many-Body Gap at Neighboring Fillings and Charge Gap} As SC is a compressible phase with broken $\mathrm{U}(1)$ symmetry, we naturally expect sufficiently small deviations in filling to preserve superconductivity. We also require the many-body ground state at such neighboring fillings to be of the same phase to compute the superconducting order parameter. Indeed, looking at the many-body gap for $N_h=N+2$ holes, we identify a single ground state at $\mathbf{K}=0$ around $d\in\left(d_{c,1},d_{c,2}\right)$~[Fig.~\ref{fig:S_AdditionalSCSignatures}(a)]. However, we observe that the many-body gap fully collapses for $N=16$~[Fig.~\ref{fig:S_AdditionalSCSignatures}(b)]. This can be attributed to the broken particle-hole symmetry, and is indeed consistent with the particle-hole transformed band exhibiting greater dispersion~[Fig.~\ref{fig:S_Bands}(c,d)]. As such, we have only compared $N_h=N$ with $N_h=N+2$ for the superconducting order parameter~[Fig.~4(a,b)].

We also consider the charge gap, which can be estimated from
\begin{equation*}
\Delta_c=\frac{E_{N+2,0}+E_{N-2,0}-2E_{N,0}}{2},
\end{equation*}
where $E_{N_h,0}$ is the ground-state energy for $N_h$ holes. We find that the charge gap is minimized in the intermediate region, consistent with the intermediate phase having broken $\mathrm{U}(1)$ charge conservation, as opposed to the incompressible fractional spin Hall insulator~[Fig.~\ref{fig:S_AdditionalSCSignatures}(c)]. Furthermore, for $N=16$, we observe a crossover in the charge gap at $d_{c,2}$ from negative to positive, consistent with the transition from the compressible superconductor to the incompressible FSHI.

\paragraph{Binding Energy and Pair Density Matrix Spectra} We note that while we include the Hartree correction when comparing energy for different magnetization sectors in [Fig.~1(b)], these are not included when computing the binding energy. This is because this correction is always positive and vanishes the thermodynamic limit, so while it may add a better estimate to the binding energy at finite sizes, it is not meaningful in distinguishing transition behavior. The binding energy $E_b^{(-)}$~[Fig.~\ref{fig:S_AdditionalSCSignatures}(d)] exhibits largely similar trends to $E_b^{(+)}$~[Fig.~3(d)].

The spectra of the superconducting pair density matrix for $N=16$~[Fig.~\ref{fig:S_AdditionalSCSignatures}(e)], similar to that of $N=14$, also has a gapped first eigenvalue peaked in the intermediate region $d\in(d_{c,1},d_{c,2})$, indicative of superconducting order. However, we note that it drops off much slower than $N=14$ beyond $d_{c,2}$. While not presented, we also note that $N=12$ follows a similar pattern to $N=12$, rather than $N=14$.

\paragraph{Superfluid Stiffness} Superfluid stiffness was computed using the finite-difference method with $\delta\Phi=0.01\Phi_0$. Let us now compare $N=14$~[Fig.~3(d)] with $N=12$~[Fig.~\ref{fig:S_AdditionalSCSignatures}(f)]. Unlike $N=14$ which saw a monotonic decrease with $d$, for $N=12$, the superfluid stiffness is maximized at an intermediate value within $\left(d_{c,1},d_{c,2}\right)$, rapidly drops off at a much faster rate around $d_{c,2}$, then attains a negative value for $d\gtrsim d_{c,2}$. The latter does not reflect upon a true instability, but is a finite-size effect from considering only one of the 36 degenerate ground states.

\paragraph{Excitonic Pair Density Matrix Spectra} We consider the equivalent of the pair density matrix, Eq.~(3), for excitonic pairs rather than Cooper pairs,
\begin{equation*}
\rho_{\mathbf{k}'\mathbf{k}}=\ev{\hat{X}_{\mathbf{k}'}^\dagger\hat{X}_{\mathbf{k}}}_0,\qquad\hat{X}_{\mathbf{k}}^\dagger=\hat{\psi}_{\uparrow,\mathbf{k}}^\dagger\hat{\psi}_{\downarrow,\mathbf{k}},
\end{equation*}
and similarly look to the spectrum as a function of $d$. We observe that for all $d>d_{c,1}$, the first eigenvalue remains close to the second eigenvalue [Fig.~\ref{fig:S_AdditionalSCSignatures}(g,h)], implying that exciton condensation fails to develop, and cannot explain the intermediate phase.

\begin{figure}[b!]
\centering
\includegraphics[width=0.99\linewidth]{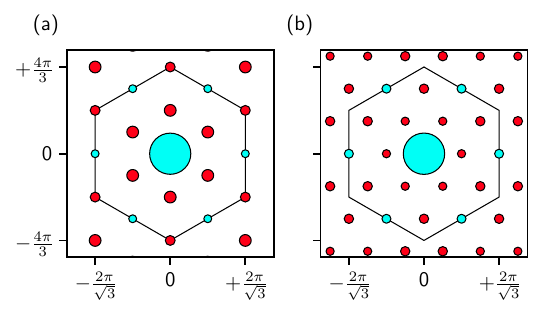}
\caption{\textbf{Maximal Pair Wavefunction.} Maximal pair wavefunction of Eq.~(3) for (a) $N=12$ and (b) $16$ at $d/a_M=0.28$.}
\label{fig:S_AdditionalPairingSymmetry}
\end{figure}

\subsection{Additional Evidence of Superconducting Pairing Symmetry}
In this subsection, we provide additional evidence for the observed superconducting pairing symmetry~[Fig.~4(a,b)].

\paragraph{Maximal Pair Wavefunction} The eigenvector of the pair density matrix, Eq.~(3), corresponding to the first eigenvalue, also corresponds to the dominant pairing channel. As such, it provides an alternative means of probing the pairing symmetry of the superconducting state. We find that the maximal pair wavefunction~[Fig.~\ref{fig:S_AdditionalPairingSymmetry}] indeed has the same pairing symmetries as the direct evaluation of the superconducting order parameter~[Fig.~4(a,b)].

\begin{figure}[b!]
\centering
\includegraphics[width=0.99\linewidth]{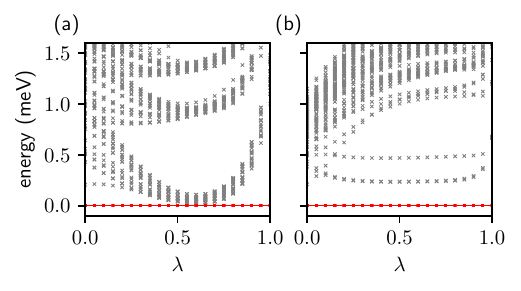}
\caption{\textbf{Adiabatic connection to ideal extended $s$-wave superconductor.} Evolution of the many-body spectrum for a linear interpolation between the Coulomb and ideal interactions for (a) $N=12$ and (b) $N=14$.}
\label{fig:S_AdiabaticConnection}
\end{figure}

\paragraph{Adiabatic Connection to Ideal Extended $s$-Wave Superconductor} We consider an ideal superconductor with the interaction,
\begin{equation*}
V^{(\text{ideal})}=-U\left(\sum_{\mathbf{k}'}f(\mathbf{k}')^*\hat{\Delta}_{\mathbf{k}'}\right)^\dagger\left(\sum_{\mathbf{k}}f(\mathbf{k})^*\hat{\Delta}_{\mathbf{k}}\right),
\end{equation*}
where $f(\mathbf{k})$ is proportional to the desired SC order parameter. As a model extended $s$-wave, we use the maximal Cooper pair wavefunction for $f(\mathbf{k})$.

For a linear interpolation between the two interactions, $V^{(\lambda)}=\left(1-\lambda\right)V^{(\text{Coulomb})}+\lambda V^{(\text{ideal})}$, where $V^{(\text{Coulomb})}$ is given by Eq.~(2), we observe that the ground state remains gapped from the excited states [Fig.~\ref{fig:S_AdiabaticConnection}].

\subsection{BCS vs. BEC}
In this subsection, we discuss whether the intermediate superconductor is better described as being the BCS limit or the BEC limit.

At $d/a_M=0.28$, the Fermi energy of the second moir\'{e} band is $E_F=1.50\:\mathrm{meV}$. When using the particle-hole transformation of the second moir\'{e} band, the Fermi energy is enhanced to $E_F'=2.70\:\mathrm{meV}$. This is comparable to the binding energy, which for $N=16$ is $|E_b| = 1.37\:\mathrm{meV}$. This suggests the intermediate superconductor lies in the BCS-BEC crossover regime. Additionally, let us compare the superconducting coherence length (Section I.f) to the lattice constant $a_M$. As noted in the main text, the short coherence length also suggests the intermediate superconductor lies in the BCS-BEC crossover regime, although direct numerical measurement of the coherence length must be carried out to make more definitive comparisons on this regard.

We note that many superconductors in moir\'{e} flat bands with non-trivial quantum geometry---magic-angle twisted bilayer graphene~\cite{oh2021evidence,tian2023evidence}, magic-angle twisted trilayer graphene~\cite{park2021tunable,kim2022evidence}, and first moir\'{e} band chiral superconductivity in MoTe${}_2$~\cite{xu2025signatures}---have been suggested to exhibit BEC--BCS crossover characteristics. These indications are based on analyses of the superconducting coherence length~\cite{tian2023evidence,park2021tunable,xu2025signatures}, the ratio $|E_b|/E_F$~\cite{oh2021evidence,kim2022evidence}, the relation between $T_{\mathrm{BKT}}$, $E_F$, and $|E_b|$~\cite{park2021tunable,oh2021evidence,cao2018unconventional}, as well as observations of pseudogap behavior~\cite{oh2021evidence,kim2022evidence}.

\section{Intermediate Superconductivity in Generalized First Landau Levels}
\paragraph{First Landau Levels} Let us define a model consisting of a first Landau level and its' time-reversal conjugate~\cite{abouelkomsan2025non}. The non-interacting energies have no dispersion, so without loss of generality, we set them as zero. For the interaction, we again use Eq.~\eqref{eq:interaction-matrix}, but replace the form factors for those of the first Landau level,
\begin{align*}
\mathfrak{f}^s_{\mathbf{k},\mathbf{P}}&=\eta_{\left[\mathbf{k}+\mathbf{P}\right]}L_1\left(\frac{1}{2}\ell_B^2\norm{\mathbf{P}}^2\right)\\
&\qquad\times e^{-\frac{1}{4}\ell_B^2\norm{\mathbf{P}}^2-\frac{i}{2}s\ell_B^2\hat{\mathbf{z}}\cdot\left(\mathbf{P}\times\mathbf{k}-\left\{\mathbf{k}+\mathbf{P}\right\}\times\left[\mathbf{k}+\mathbf{P}\right]\right)},
\end{align*}
where $L_1(x)$ is the first Laguerre polynomial and $\eta_\bullet\::\:\hat{\Lambda}\rightarrow\{\pm 1\}$ is a sign factor defined by taking some representation of the reciprocal lattice $\hat{\Lambda}=\mathrm{span}_{\mathbb{Z}}\left(\mathbf{q}_1,\mathbf{q}_2\right)$ and setting $\eta_{\mathsf{l}_1\mathbf{q}_1+\mathsf{l}_2\mathbf{q}_2}=\left(-1\right)^{\mathsf{l}_1+\mathsf{l}_2+\mathsf{l}_1\mathsf{l}_2}$, i.e. only $+1$ when $\mathsf{l}_1$ and $\mathsf{l}_2$ are both even, and $-1$ otherwise.

\paragraph{Generalized First Landau Levels} We use a generalization of first Landau level wavefunctions given by
\begin{equation*}
\psi_{\mathbf{k}}(\mathbf{r})=\mathcal{N}_\mathbf{k}e^{\phi(\mathbf{r})}\psi^{\left(\text{1LL}\right)}_{\mathbf{k}}(\mathbf{r}),
\end{equation*}
where $\mathcal{N}_{\mathbf{k}}$ is a normalization factor, and $\phi\left(\mathbf{r}\right)$ is a lowest harmonic modulation which preserves the lattice symmetries,
\begin{equation*}
\phi(\mathbf{r})=\frac{\sqrt{3}}{4\pi}K\sum_{i}\cos\left(\mathbf{Q}_i\cdot\mathbf{r}\right),
\end{equation*}
with a parameter $K$ which controls the strength of the modulation.

We note that a similar generalization for LLLs has been often used in context of quantum geometry to introduce non-uniformity while maintaining ideality~\cite{wang2021exact,fujimoto2025higher,liu2025theory}, and was recently used to stabilize chiral superconductivity proximate to the $\nu_h=\frac{2}{3}$ fractional quantum anomalous Hall state~\cite{guerci2025fractionalization}.

\begin{figure}[b!]
\centering
\includegraphics{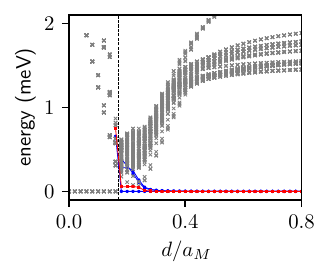}
\caption{\textbf{Evolution of many-body spectra for the ideal first Landau level and its time reversal conjugate for $N=14$.} The states connecting to the ground states of the fractional spin Hall insulator in the $d\to\infty$ limit are colored (red square or blue circle). We do not observe the intermediate state with a singular gapped state with total momentum $\mathrm{K}=0$ (red squares), which is instead, immediately degenerate with the state with total momentum $\mathrm{K}=7$.}
\label{fig:S_1LLs}
\end{figure}

\begin{figure}[b!]
\centering
\includegraphics[width=0.99\linewidth]{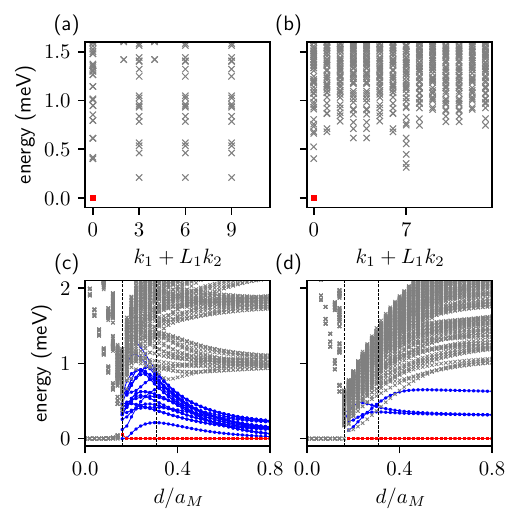}
\caption{\textbf{Many-body spectra for a generalized first Landau level and its time reversal conjugate.} (a,b) Many-body spectrum at $d=0.28$ for $N=12$ and $14$, respectively. (c,d) Evolution of the many-body spectra over $d$ for $N=12$ and $14$, respectively.}
\label{fig:S_g1LLs}
\end{figure}

\paragraph{Many-Body Spectra} When considering exact first Landau levels ($K=0$), for $N=14$, we observe that the many-body spectrum evolves directly from the ferromagnetic Chern insulator to the fractional spin Hall insulator~[Fig.~\ref{fig:S_1LLs}]. Introducing non-uniform quantum geometry with $K=0.1$, the many-body spectrum exhibits the singular ground state at $d=0.28$~[Fig.~\ref{fig:S_g1LLs}(a,b)], and the overall evolution of the spectra become highly similar to that of the continuum model of twisted bilayer MoTe$_2$~[Fig.~\ref{fig:S_g1LLs}(c,d)]. This suggests non-uniform quantum geometry is a key ingredient in stabilizing the intermediate superconductor.

\section{Kohn-Luttinger Mechanism for the Intermediate Superconductor}
\paragraph{Random Phase Approximation} Here we present the random phase approximation, incorporating quantum geometry and spin anisotropy. The screened interaction is given by
\begin{align*}
&\overline{V}^{ss'}\left(\mathbf{p};\mathbf{Q},\mathbf{Q}'\right)\\
&=\sum_{\tilde{s}}\left[\left(\mathtt{I}+\mathtt{Z}_{\mathbf{p}}\right)^{-1}\right]_{\left(s,\mathbf{Q}\right),\left(\tilde{s},\mathbf{Q}'\right)}V^{\tilde{s}s'}(\mathbf{p}+\mathbf{Q}'),
\end{align*}
where $\mathtt{I}$ and $\mathtt{Z}_{\mathbf{p}}$ are dimensionless matrices with indices in $\mathbb{Z}_2^\text{spin}\times\hat{\Lambda}$. $\mathtt{I}$ is the identity matrix, and $\mathtt{Z}_{\mathbf{p}}$ is given by
\begin{equation*}
\left[\mathtt{Z}_{\mathbf{p}}\right]_{\left(s,\mathbf{Q}\right),\left(s',\mathbf{Q}'\right)}=V^{ss'}(\mathbf{p}+\mathbf{Q})\chi^{s'}_{\mathbf{p},\mathbf{Q}\mathbf{Q}'},
\end{equation*}
and $\chi^s_{\mathbf{p},\mathbf{Q}\mathbf{Q}'}$ is the quantum geometry weighted static polarization,
\begin{equation*}
\chi^s_{\mathbf{p},\mathbf{Q}\mathbf{Q}'}=\frac{1}{A}\sum_{\mathbf{k}}\overline{\mathfrak{f}}^s_{\mathbf{k},\mathbf{p}+\mathbf{Q}}\mathfrak{f}^s_{\mathbf{k},\mathbf{p}+\mathbf{Q}'}\tau^s_{\mathbf{k},\mathbf{p}},
\end{equation*}
where
\begin{equation*}
\tau^s_{\mathbf{k},\mathbf{p}}=-\frac{\left(1+e^{\left(\epsilon^s_{\{\mathbf{k}+\mathbf{p}\}}-\mu\right)/T}\right)^{-1}-\left(1+e^{\left(\epsilon^s_{\mathbf{k}}-\mu\right)/T}\right)^{-1}}{\epsilon^s_{\{\mathbf{k}+\mathbf{p}\}}-\epsilon^s_{\mathbf{k}}}.
\end{equation*}
Note that due to the ambiguity of momentum up to reciprocal lattice momentum under the creation and annihilation of electron-hole pairs, it is possible $\mathbf{Q}\neq\mathbf{Q}'$.

In twisted bilayer MoTe${}_2$ second moir\'{e} bands, the first Landau level-like quantum geometry implies the form factors decay exponentially at the rate of the magnetic length. As such, the static polarization can be reasonably approximated as diagonal along the reciprocal lattice index,
\begin{equation*}
\chi^s_{\mathbf{p},\mathbf{Q}\mathbf{Q}'}\approx\delta_{\mathbf{Q}\mathbf{Q}'}\chi^s_{\mathbf{p}+\mathbf{Q}},
\end{equation*}
where
\begin{equation*}
\chi^s_{\mathbf{P}}=\frac{1}{A}\sum_{\mathbf{k}}\left|\mathfrak{f}^s_{\mathbf{k},\mathbf{P}}\right|^2\tau^s_{\mathbf{k},\{\mathbf{P}\}}.
\end{equation*}
The screened interaction is then expressed as a potential,
\begin{equation}
\overline{V}^{ss'}\left(\mathbf{P}\right)=\sum_{\tilde{s}}\left[\left(\mathtt{I}+\mathtt{Z}_{\mathbf{P}}\right)^{-1}\right]^{s\tilde{s}}V^{\tilde{s}s'}(\mathbf{P}),
\label{eq:RPA-screening}
\end{equation}
where $\mathtt{I}$ is the $2\times 2$ identity matrix over spin and 
\begin{equation*}
\left[\mathtt{Z}_{\mathbf{P}}\right]^{ss'}=V^{ss'}(\mathbf{P})\chi^{s'}_{\mathbf{P}}.
\end{equation*}
We use this expression for our calculations.

\begin{figure}[b!]
\centering
\includegraphics[width=0.99\linewidth]{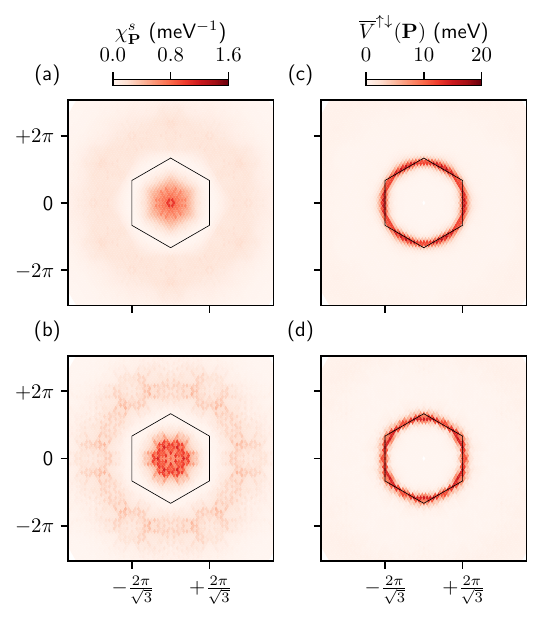}
\caption{\textbf{RPA Screened Intervalley Interaction.} (a,b) Static polarization using (a) the second moir\'{e} band at $\nu_h=2$, (b) and the particle-hole transformed band, for $d/a_M=0.28$. Due to time-reversal and valley-reduced inversion symmetry, the static polarizations are the same for both valleys. (c,d) The respective RPA screened intervalley interaction potential.}
\label{fig:S_RPAScreening}
\end{figure}

\paragraph{RPA Screening in tbMoTe${}_2$ Second Moir\'{e} Bands} We have calculated the RPA-screened intervalley interaction for the second moir\'{e} bands at $d/a_M=0.28$~[Fig.~4(c)]. Here we elaborate on various contributing factors listed in the main text.

As discussed in Section I.d, because the second moir\'{e} band has non-uniform quantum geometry and is very flat relative to the interaction strength, it is non-trivially renormalized under particle-hole transformation~[Fig.~\ref{fig:S_Bands}(c,d)]. In addition to increasing the bandwidth, while the original band has only a single Fermi surface pocket, the particle-hole transformed band has three Fermi surface pockets. The presence of multiple Fermi surface pockets have been argued to be beneficial in stabilizing Kohn-Luttinger superconductivity~\cite{maiti2010renormalization}, and is likely relevant for $\nu_h\gtrsim 3$. Additionally, the second moir\'{e} band hosts van Hove singularities at the $\mu$ points, which are close to the $\nu_h=3$ Fermi surface, which even change occupation of the $\mu$ points under particle-hole transformation~[Fig.~\ref{fig:S_Bands}(c,d)]. As the van Hove singularities have divergent density of states, they provide large leading contribution to the static polarization, and thus, significant RPA screening~\cite{gonzalez2008kohn,nandkishore2012chiral}.

Let us now discuss the static polarization, $\chi_{\mathbf{P}}^s$~[Fig.~\ref{fig:S_RPAScreening}(a,b)]. We note that both spins give the same expression for the static polarization. For both the electron and hole basis, we observe similar overall radial features in $\chi_{\mathbf{P}}^s$. It is largest at small $\norm{\mathbf{P}}$, quickly decreases in magnitude and attains a minima at the edges of the first Brillouin zone, $\norm{\mathbf{P}}\sim Q/2$ where $Q=\norm{\mathbf{Q}_1}$, then increases again to attain a much smaller maxima at $\norm{\mathbf{P}}\sim Q$, and so forth. There are also non-trivial $C_3^z$-symmetric angular features, which are much more pronounced for the particle-hole transformed band. The rapid decay of the static polarization with $\mathbf{P}$ can primarily be attributed to the first Landau level like character of the second moir\'{e} band, while the non-trivial radial and angular features result from non-uniform quantum geometry and Fermi surface pockets. In particular, it has been noted that if the quantum geometry is uniform, or more generally, if the quantum distance along the Fermi surface is in some sense equal to the quantum geometry-driven under-screening driven by the form factors between the Fermi sea and complementary unoccupied states, the two effects can cancel out, resulting in a monotonically decreasing static polarization. This occurs in exactly uniform setups such as spin-polarized Landau levels~\cite{shavit2025quantum}.

Substituting the static polarization into Eq.~\eqref{eq:RPA-screening}, we obtain the RPA screened intervalley interaction~[Fig.~\ref{fig:S_RPAScreening}(c,d)]. We note that due to spin anisotropy of the bare interaction, Eq.~(2), the intervalley interaction is more strongly screened than the intravalley interaction. Unlike the bare intervalley interaction, which monotonically decreases with $\norm{\mathbf{P}}$, the RPA screened intervalley interaction \textit{increases} in momentum within the first Brillouin zone, attaining a maxima along its' edges $\norm{\mathbf{P}}\approx Q/2$, before dropping off. This is true for both the electron and hole basis. Such behavior is highly desirable for stabilizing Kohn-Luttinger superconductivity, and typically leads to nodal superconductivity~\cite{shavit2025quantum}. As such, the non-uniform quantum geometry is a critical ingredient for the Kohn-Luttinger mechanism in quantum geometry driven superconductivity. The two bases differ in the $C_3^z$ angular features, with the particle-hole transformed band exhibiting a greater amplitude to the $C_3^z$ angular features, likely further stabilizing Kohn-Luttinger superconductivity.

\paragraph{Linearized BCS Gap Equation and Kohn-Luttinger Pairing Instability} At the mean-field level, the superconducting gap satisfies the BCS self-consistent gap equation,
\begin{equation*}
\Delta_{\mathbf{k}}=-\frac{1}{2A}\sum_{\mathbf{k}'}\mathfrak{V}_{\mathbf{k}\mathbf{k}'}\frac{\tanh\left(\xi_{\mathbf{k}'}/2T\right)}{\xi_{\mathbf{k}'}}\Delta_{\mathbf{k}'},
\end{equation*}
where $\mathfrak{V}_{\mathbf{k}\mathbf{k}'}$ is the reduced interaction matrix,
\begin{equation*}
\mathfrak{V}_{\mathbf{k}\mathbf{k}'}=\sum_{\mathbf{Q}}\left|\mathfrak{f}^\uparrow_{\mathbf{k}',\mathbf{k}-\mathbf{k}'+\mathbf{Q}}\right|^2\tilde{V}^{\uparrow\downarrow}\left(\mathbf{k}-\mathbf{k}'+\mathbf{Q}\right).
\end{equation*}
and $\xi_{\mathbf{k}}=\sqrt{\left(\epsilon_{\mathbf{k}}-\mu\right)^2+\Delta_{\mathbf{k}}^2}$ gives the energy of the Bogoliubov quasiparticles. We solved the linearized BCS gap equation for the RPA screened intervalley interaction to determine the pairing instability~[Fig.~4(d)].

\paragraph{Limitations of Kohn-Luttinger Analysis} We note that, although we obtain a nodal extended $s$-wave pairing instability in the $A_1$ irreducible representation---consistent with the pairing symmetry found in the exact diagonalization calculations---some discrepancies remain, such as the sign reversal of the pairing function between the $\gamma$ and $\mu$ points. We attribute such deviations to the limitations of the Kohn–Luttinger pairing instability, which is derived within a mean-field and perturbative framework, whereas in our system the Coulomb interaction exceeds the bandwidth and the superconducting state resides in the BCS--BEC crossover regime (Section~II.F), where strong interactions beyond perturbative treatments become important.

\section{Field-Theoretical Description of the FSHI-SC Transition}
In this section, we provide additional details on our field-theoretical description of the transition from the fractional spin Hall insulator to the superconductor.

\paragraph{Effective Field Theory of the Fractional Spin Hall Insulator} The fractional spin Hall insulator can be viewed as a time-reversal symmetric pair of the Pfaffian states, i.e., $\mathrm{Pf} \times \overline{\mathrm{Pf}}$. The effective field theory of the Pfaffian state is \cite{fradkin1998chern}
\begin{align*}
\mathcal{L}_\mathrm{Pf} &= - \frac{2}{4 \pi} \Tr\left(a \diff a + \frac{2}{3} a^3\right) + \frac{3}{4 \pi} (\Tr a) \diff (\Tr a) \nonumber \\
&\quad + \frac{1}{2 \pi} A \diff \Tr a + \frac{1}{4 \pi} A \diff A + 2 \mathrm{CS}_g,
\end{align*}
where $a$ is a dynamical $\mathrm{U}(2)$ gauge field, $A$ is the background electromagnetic field, and $\mathrm{CS}_g$ is the gravitational Chern-Simons term. Its time-reversal conjugate is given by 
\begin{align*}
\mathcal{L}_{\overline{\mathrm{Pf}}} &= \frac{2}{4 \pi} \Tr\left(a \diff a + \frac{2}{3} a^3\right) - \frac{3}{4 \pi} (\Tr a) \diff (\Tr a) \nonumber \\
&\quad - \frac{1}{2 \pi} A \diff \Tr a - \frac{1}{4 \pi} A \diff A - 2 \mathrm{CS}_g.
\end{align*}
Thus, the total Lagrangian describing the fractional spin Hall insulator is 
\begin{align} \label{eq:fqsh_lagrangian}
\mathcal{L}_\mathrm{FSHI} &= - \frac{2}{4 \pi} \Tr\left(a \diff a + \frac{2}{3} a^3\right) + \frac{3}{4 \pi} (\Tr a) \diff (\Tr a) \nonumber \\
&\quad + \frac{2}{4 \pi} \Tr\left(b \diff b + \frac{2}{3} b^3\right) - \frac{3}{4 \pi} (\Tr b) \diff (\Tr b) \nonumber \\
&\quad + \frac{1}{2 \pi} A \diff (\Tr a - \Tr b),
\end{align}
where $b$ is another dynamical $\mathrm{U}(2)$ gauge field. The Lagrangian Eq.~\eqref{eq:fqsh_lagrangian} describes the topological order of the fractional spin Hall insulator, which is given by $\mathrm{Pf} \times \overline{\mathrm{Pf}}$.

\paragraph{Anyon Condensation Transition} The transition to the superconductor is induced by the condensation of a complex scalar field $\Phi$ that transforms as the bifundamental representation of $\mathrm{U}(2) \times \mathrm{U}(2)$, whose covariant derivative is given by 
\begin{align*}
D_\mu \Phi = \partial_\mu \Phi - i \sum_{i = 0}^3 \left(a_\mu^l T^l \Phi + b_\mu^l \Phi T^l\right),
\end{align*}
where $T^0 = I / 2$ and $T^l = \sigma^l / 2$ are the generators of the Lie algebra $\mathfrak{u}(2)$. The scalar field $\Phi$ is identified as the composite of the charge-$\frac{1}{4}$ non-Abelian anyon $\sigma_{\frac{1}{4}}$ in the Pfaffian state $\mathrm{Pf}$ and its time-reversed partner $\sigma_{\frac{1}{4}}^\mathcal{T}$ in the time-reversed Pfaffian state $\overline{\mathrm{Pf}}$. Anyons in $\mathrm{U}(2)$ Chern-Simons theory are labeled by $(j, n)$ with $j + \frac{n}{2} \in \mathbb{Z}$, where $2j$ and $n$ are integers~\cite{shi2025doping_}. Here, $j$ and $n$ correspond to an irreducible representation of $\mathrm{SU}(2)$ and a charge of $\mathrm{U}(1)$, respectively. In this framework, $\Phi$ corresponds to the composite of the anyons labeled $(\frac{1}{2}, 1)$ in $\mathrm{Pf}$ and $(\frac{1}{2}, 1)$ in $\overline{\mathrm{Pf}}$. The anyon $(\frac{1}{2}, 1)$ in $\mathrm{Pf}$ can be identified with the charge-$\frac{1}{4}$ non-Abelian anyon $\sigma_{\frac{1}{4}}$. To further demonstrate consistency, we check the electric charge of the anyon. The Abelian sector of $\mathrm{Pf}$, associated with the gauge field $a$, is described by a $\mathrm{U}(1)_8$ theory. Including the coupling to the external electromagnetic gauge field $A$, this Abelian sector can be represented by the $K$-matrix $K = (8)$ and the charge vector $\mathbf{t} = (2)$. The electric charge of an anyon labeled by integer vector $\mathbf{l} = (1)$ is then given by $q_\mathbf{l} = \mathbf{t}^\mathsf{T} K^{-1} \mathbf{l} = \frac{1}{4}$.

When the gauge-invariant operator $\Tr \Phi^\dagger \Phi$ acquires a finite vacuum expectation value, i.e., $\langle \Tr \Phi^\dagger \Phi \rangle \neq 0$, the $\mathrm{U}(2) \times \mathrm{U}(2)$ gauge group is spontaneously broken. To understand how the condensation reduces the gauge group, we denote the singular values of $\Phi$ by $s_1$ and $s_2$. Expressed in terms of these singular values, the potential energy $V(\Phi^\dagger \Phi)$ in our Chern-Simons-Landau-Ginzburg theory takes the form 
\begin{align*}
V &= m^2 \Tr \Phi^\dagger \Phi + \lambda_1 \Tr \Phi^\dagger \Phi \Phi^\dagger \Phi + \lambda_2 (\Tr \Phi^\dagger \Phi)^2 \nonumber \\
&= m^2 (s_1^2 + s_2^2) + \frac{\lambda_1 + \lambda_2}{2} (s_1^2 + s_2^2)^2 + \frac{\lambda_1}{2} (s_1^2 - s_2^2)^2.
\end{align*}
For our purposes, we take $m^2 < 0$ to ensure condensation, and impose $\lambda_1 > 0$ and $\lambda_1 + \lambda_2 > 0$ to guarantee that the potential energy is bounded from below. The potential energy $V(\Phi^\dagger \Phi)$ is minimized when the singular values are equal as $s \equiv s_1 = s_2$, with 
\begin{align*}
s = \sqrt{\frac{- m^2}{2\left(\lambda_1 + \lambda_2\right)}}.
\end{align*}
Choosing a gauge in which the condensate takes the form $\langle \Phi \rangle = s I$ with $s > 0$, the kinetic term becomes 
\begin{align*}
\Tr\left(D^\mu \langle \Phi \rangle^\dagger D_\mu \langle \Phi \rangle\right) = \frac{s^2}{2} \sum_{l = 0}^3 (a_\mu^l + b_\mu^l)^2.
\end{align*}
This shows that, in the condensed phase, the combination $a_\mu + b_\mu$ acquires a mass, effectively imposing the constraint $\alpha \equiv a = - b$ at low energies. Substituting this into the Lagrangian Eq.~\eqref{eq:fqsh_lagrangian}, we obtain the effective field theory describing the superconductor~\cite{hansson2004superconductors},
\begin{align} \label{eq:sc_lagrangian}
\mathcal{L}_\mathrm{SC} = \frac{2}{2 \pi} A \diff\Tr \alpha.
\end{align}
Transitions of this type have been extensively studied in literature and are widely believed to be continuous~\cite{wen1993transitions,kou2009mutual,barkeshli2010anyon,barkeshli2014continuous,han2025anyon,zhou2025chern,huang2025third,ji2026self,wu2023continuous,gazit2018confinement,zeng2020continuous,schuler2023emergent,hwang2024anyon,wang2026emergent}, supported by both field-theoretical arguments~\cite{wen1993transitions,kou2009mutual,barkeshli2010anyon,barkeshli2014continuous,han2025anyon,zhou2025chern,huang2025third,ji2026self,wu2023continuous} and numerical studies~\cite{wu2023continuous,gazit2018confinement,zeng2020continuous,schuler2023emergent,hwang2024anyon,wang2026emergent}.

\paragraph{Effective Field Theory of the Superconductor} We now discuss Eq.~\eqref{eq:sc_lagrangian} in more detail, showing that it indeed describes a charge-$2e$ superconductor. Consider an arbitrary closed $2$-manifold $\Sigma_2$ and cover it with two open patches $N$ and $S$, which overlap at the ``equator'' $E \simeq S^1$. In the overlap region $E$, the locally defined $\mathrm{U}(2)$ gauge fields $\alpha_N$ and $\alpha_S$ are related by a gauge transformation 
\begin{align*}
\alpha_N = g^{-1} \alpha_S g - i g^{-1}\diff g,
\end{align*}
where $g: E \to \mathrm{U}(2)$. Taking the trace, we get 
\begin{align*}
\Tr \alpha_N = \Tr \alpha_S - i\diff(\log \det g),
\end{align*}
where we have used the identity $\Tr g^{-1}\diff g = \diff(\log \det g)$. Then, by integrating $\diff\Tr \alpha$ over $\Sigma_2$, we get 
\begin{align*}
\int_{\Sigma_2} \diff\Tr \alpha = - \int_E\diff(\log \det g).
\end{align*}
Note that $\det g$ maps $E \simeq S^1$ into $\mathrm{U}(1)$. Since the integral of the derivative of the logarithm of a map $S^1 \to \mathrm{U}(1)$ is given by $2 \pi i$ times the winding number of the map, we have 
\begin{align*}
\int_{\Sigma_2}\diff\Tr \alpha \in 2 \pi \mathbb{Z}.
\end{align*}
Thus, the flux of $\Tr \alpha$ is quantized in $2 \pi$. Since there is no Chern-Simons term, the $2 \pi$ monopole of $\Tr \alpha$, which is bound to the $\mathrm{SU}(2)$ $\pi$-flux, is gauge-invariant, and can condense and lead to the confinement. In the absence of gapless matter fields coupled to the gauge fields, and in the absence of topological terms (recall that $A$ is a nondynamical background gauge field and thus the mutual Chern–Simons coupling to $A$ does not generate a mass term for the dynamical gauge fields), the gauge field $\alpha$ is expected to be confining~\cite{polyakov1977quark,teper1993confining}, corresponding to the superconducting state of interest. Since the mutual Chern-Simons level between $A$ and $\Tr \alpha$ is $2$, the condensation of the monopole induces charge-$2e$ superconductivity.

\paragraph{Additional Details of the Condensation Transition} To complete the analysis, we will show that all $8$ components of $\Phi$ are either eaten by the massive vector fields or gapped. To see which components are eaten, we express $\Phi$ as 
\begin{align*}
\Phi = e^{i \theta^a T^a / s} (s I + H)
\end{align*}
using the polar decomposition. Here, $\theta^a$ are $4$ real phase fields and $H$ is a $2 \times 2$ Hermitian matrix. Note that $H$ also contains $4$ real scalar fields, so that the total degrees of freedom of $\Phi$ is $8$ as expected. Since every unitary matrix can be written as the square of another unitary matrix, the scalar field can be written as a Hermitian matrix via a gauge transformation. Note that $4$ degrees of freedom $\theta^a$ have been absorbed into the redefinition of the gauge fields, or ``eaten'' by the gauge fields. With this gauge choice, the scalar field can be written as 
\begin{align*}
\Phi = 
\begin{pmatrix}
s + \sigma_1 & x - i y \\
x + i y & s + \sigma_2
\end{pmatrix},
\end{align*}
where $\sigma_j$, $x$, and $y$ are real scalar fields. Note that these four fields correspond to the remaining uneaten components of $\Phi$. By plugging this into the potential energy, we get 
\begin{align*}
V &= 4 s^2 (\lambda_1 - s(\lambda_1 + \lambda_2)) (\sigma_1 + \sigma_2) \nonumber \\
&\quad + 2 s^2 (2 \lambda_1 + \lambda_2) (\sigma_1^2 + \sigma_2^2) \nonumber \\
&\quad + 8 s^2 \lambda_1 (x^2 + y^2) + \cdots,
\end{align*}
where $\cdots$ includes higher-order terms. We find that all four real scalar fields acquire a finite mass. Specifically, the mass of $\sigma_j$ are $2 s \sqrt{2 \lambda_1 + \lambda_2}$, while the mass of $x$ and $y$ are $4 s \sqrt{\lambda_1}$. Thus, in the effective Lagrangian, these components are gapped out and do not appear.

\section{Categorical Description of the FSHI-SC Transition}
In this section, we supplement our field-theoretical argument, which is on its' own sufficient and self-contained, with a category-theoretic formulation of anyon superconductivity developed by some of our authors~\cite{seo2026unified}, and show that the results are consistent. Within this framework, anyon superconductivity is understood as a generalized anyon condensation that explicitly keeps track of charge conservation, and the breaking of charge conservation.

\paragraph{General Categorical Description of Anyon-Driven Superconductivity} Topological orders in $2 + 1$ dimensions host exotic quasiparticles known as anyons and the anyon theories are classified by modular tensor categories~\cite{wen2015theory,bruillard2017fermionic}. When a given topological order has global $\mathrm{U}(1)$ charge conservation symmetry, the anyons carry fractionalized electric charges~\cite{barkeshli2019symmetry}. Although the fractional charges of the anyons are captured modulo $e$ in the standard modular tensor category formalism, to fully describe the transition into a superconducting phase via anyon condensation, it is necessary to track the exact values of the anyon charges. 

To do so, modular tensor categories \textit{over} $\mathrm{Rep}(\mathrm{U}(1))$ or $\mathrm{sRep}(\mathrm{U}(1)^f)$ are introduced in Ref.~\cite{seo2026unified}. The subcategories $\mathrm{Rep}(\mathrm{U}(1))$ and $\mathrm{sRep}(\mathrm{U}(1)^f)$ describes the local excitations, i.e., those do not exhibit anyonic statistics, carrying integer-valued electric charges in units of $e$. The fractional charges of the anyons are determined by their fusion rules involving these local excitations. Then, the transition into a superconductor is described by a condensable algebra $\mathcal{A}$, which represents condensing anyons during the transition. When the anyons in $\mathcal{A}$ are charged, then $\mathcal{A}$ must contain charged local excitations since $\mathcal{A}$ includes at least one fusion channel of its constituent anyons. In consequence, the condensation of $\mathcal{A}$ breaks $\mathrm{Rep}(\mathrm{U}(1))$ or $\mathrm{sRep}(\mathrm{U}(1)^f)$ down to a smaller subcategory. Physically, this means that the global $\mathrm{U}(1)$ charge conservation symmetry is broken down to a subgroup, implying that a superconducting phase emerges. The charge quantization of the resulting superconductor is determined by the smallest local charge included in $\mathcal{A}$.

More precisely, the category $\mathrm{Rep}(\mathrm{U}(1))$ consists of a countably infinite number of simple objects, labeled by integers $\cdots, \Bar{\mathbf{2}}, \Bar{\mathbf{1}}, \mathbf{0}, \mathbf{1}, \mathbf{2}, \cdots$. Physically, the integers correspond to conserved $\mathrm{U}(1)$ charges of local particle-like excitations. All the simple objects braid trivially with one another and have bosonic self-statistics. While the category $\mathrm{sRep}(\mathrm{U}(1)^f)$ also consists of integer-labeled simple objects, the odd-integer objects now have fermionic self-statistics. This can be viewed as a categorical implementation of the spin-charge relation~\cite{seiberg2016gapped}, which is obeyed in ordinary condensed matter composed of electrons. When the condensable algebra $\mathcal{A}$ contains a local boson $\mathbf{m}$, the condensation of $\mathcal{A}$ imposes the equivalence relation $\mathbf{n} \sim \mathbf{n} + \mathbf{m}$ to the integer-labeled objects. In consequence, $\mathrm{U}(1)$ charges of the local excitations are defined only modulo $\mathbf{m}$ in the condensed phase. Categorically, $\mathrm{Rep}(\mathrm{U}(1))$ (or $\mathrm{sRep}(\mathrm{U}(1)^f)$) is mapped into $\mathrm{Rep}(\mathbb{Z}_m)$ (or $\mathrm{sRep}(\mathbb{Z}_m^f)$) due to the condensation of $\mathcal{A}$, which describe onsite $\mathbb{Z}_m$ symmetries~\cite{lan2017classification}. Thus, the above process can be viewed as a categorical description of symmetry breaking $\mathrm{U}(1)$ down to $\mathbb{Z}_m$, which induces charge-$m e$ superconductivity~\cite{seo2026unified}.

\paragraph{Categorical Description of the FSHI-SC Transition} Now, let us apply the above categorical formalism to the transition from the fractional spin Hall state $\mathrm{Pf} \times \overline{\mathrm{Pf}}$ to the charge-$2e$ superconductor. We first recall that the Pfaffian state contains six distinct anyon types, labeled by $\mathbf{0}$, $\sigma_{\frac{1}{4}}$, $\psi_0$, $\alpha_{\frac{1}{2}}$, $\Bar{\sigma}_{\frac{3}{4}}$, and $\Bar{\alpha}_{\frac{1}{2}}$, where the subscripts denote their electric charges in units of the electron charge $e$ (with $\mathbf{0}$ representing the vacuum sector)~\cite{bruillard2017fermionic}. The mutual statistics of the anyons are encoded in the (fermion-quotient) $S$ matrix \cite{bruillard2017fermionic}:
\begin{align*}
    S = \frac{1}{2 \sqrt{2}}
    \begin{pmatrix}
        1 & \sqrt{2} & 1 & 1 & \sqrt{2} & 1 \\
        \sqrt{2} & 0 & -\sqrt{2} & i \sqrt{2} & 0 & -i \sqrt{2} \\
        1 & -\sqrt{2} & 1 & 1 & -\sqrt{2} & 1 \\
        1 & i \sqrt{2} & 1 & -1 & -i \sqrt{2} & -1 \\
        \sqrt{2} & 0 & -\sqrt{2} & -i \sqrt{2} & 0 & i \sqrt{2} \\
        1 & -i \sqrt{2} & 1 & -1 & i \sqrt{2} & -1
    \end{pmatrix}.
\end{align*}
More precisely, the mutual braiding phase of a pair of anyons $i$ and $j$ are given by $S_{ij}^* S_{11} / S_{1i} S_{1j}$ \cite{barkeshli2019symmetry}. In a $\mathrm{U}(2)$ gauge theory~\cite{shi2025doping_}, these anyons correspond respectively to $(k = 0, n = 0)$, $(\frac{1}{2}, 1)$, $(1, 0)$, $(0,2)$, $(\frac{1}{2}, 3)$, and $(1, 2)$. The anyon content of the time-reversal-conjugate consists of the corresponding time-reversed partners: $\mathbf{0}$, $\sigma_{\frac{1}{4}}^\mathcal{T}$, $\psi_0^\mathcal{T}$, $\alpha_{\frac{1}{2}}^\mathcal{T}$, $\Bar{\sigma}_{\frac{3}{4}}^\mathcal{T}$, and $\Bar{\alpha}_{\frac{1}{2}}^\mathcal{T}$. The mutual braiding statistics of the time-reversed anyons can be simply obtained by taking complex conjugation on the mutual statistics of the anyons in the Pfaffian theory. Since we are considering a bilayer system where each layer host the Pfaffian theory and its time-reversal, any pair of anyons consisting of one from the Pfaffian and the other from the time-reversed Pfaffian always braid trivially. Therefore, the full anyon content of the fractional quantum spin Hall state, described by $\mathrm{Pf} \times \overline{\mathrm{Pf}}$, consists of all composite pairs formed by combining anyons from $\mathrm{Pf}$ and those from $\overline{\mathrm{Pf}}$. 

The transition described by the field-theoretical framework above corresponds to an anyon condensation characterized by the condensable algebra
\begin{align*}
\mathcal{A} &= \bigoplus_{\mathbf{m} \in 2 \mathbb{Z}} \mathbf{m} \nonumber \\
&\quad \cdot \left(\mathbf{0} \oplus \psi_0 \psi_0^\mathcal{T} \oplus \sigma_{\frac{1}{4}} \sigma_{\frac{1}{4}}^\mathcal{T} \oplus \alpha_{\frac{1}{2}} \alpha_{\frac{1}{2}}^\mathcal{T} \oplus \Bar{\alpha}_{\frac{1}{2}} \Bar{\alpha}_{\frac{1}{2}}^\mathcal{T} \oplus \Bar{\sigma}_{\frac{3}{4}} \Bar{\sigma}_{\frac{3}{4}}^\mathcal{T}\right).
\end{align*}
Note that the fusion rules of the anyons are given by
\begin{align*}
&\sigma_{\frac{1}{4}} \otimes \sigma_{\frac{1}{4}} = \alpha_{\frac{1}{2}} \oplus \Bar{\alpha}_{\frac{1}{2}}, \nonumber \\ &\sigma_{\frac{1}{4}} \otimes \alpha_{\frac{1}{2}} = \sigma_{\frac{1}{4}} \otimes \Bar{\alpha}_{\frac{1}{2}} = \Bar{\sigma}_{\frac{3}{4}}, \nonumber \\
&\sigma_{\frac{1}{4}} \otimes \Bar{\sigma}_{\frac{3}{4}} = \mathbf{1} \oplus \mathbf{1} \psi_0, \\
&\alpha_{\frac{1}{2}} \otimes \alpha_{\frac{1}{2}} = \Bar{\alpha}_{\frac{1}{2}} \otimes \Bar{\alpha}_{\frac{1}{2}} \mathbf{1} \psi_0, \\
&\alpha_{\frac{1}{2}} \otimes \Bar{\alpha}_{\frac{1}{2}} = \mathbf{1}, \\
&\alpha_{\frac{1}{2}} \otimes \Bar{\sigma}_{\frac{3}{4}} = \Bar{\alpha}_{\frac{1}{2}} \otimes \Bar{\sigma}_{\frac{3}{4}} = \mathbf{1} \sigma_{\frac{1}{4}}.
\end{align*}
The fusion rules of the time-reversed anyons with the superscript $\mathcal{T}$ are given as the same. Therefore, all the anyons in $\mathcal{A}$ appear in the repeated fusion of $\sigma_{\frac{1}{4}} \sigma_{\frac{1}{4}}^\mathcal{T}$. For example, 
\begin{align*}
(\sigma_{\frac{1}{4}} \sigma_{\frac{1}{4}}^\mathcal{T})^2 &\supset \alpha_{\frac{1}{2}} \alpha_{\frac{1}{2}}^\mathcal{T} \oplus \Bar{\alpha}_{\frac{1}{2}} \Bar{\alpha}_{\frac{1}{2}}^\mathcal{T}, \nonumber \\
(\sigma_{\frac{1}{4}} \sigma_{\frac{1}{4}}^\mathcal{T})^3 &\supset \Bar{\sigma}_{\frac{3}{4}} \Bar{\sigma}_{\frac{3}{4}}^\mathcal{T}, \nonumber \\
(\sigma_{\frac{1}{4}} \sigma_{\frac{1}{4}}^\mathcal{T})^4 &\supset \mathbf{2} \oplus \mathbf{2} \psi_0, \nonumber \\
(\sigma_{\frac{1}{4}} \sigma_{\frac{1}{4}}^\mathcal{T})^4 \otimes \Bar{\mathbf{2}} &\supset \mathbf{0} \oplus \psi_0.
\end{align*}
Since $\mathcal{A}$ contains the charge-$2e$ local boson $\mathbf{2}$, the condensation of $\mathcal{A}$ breaks $\mathrm{sRep}(\mathrm{U}(1)^f)$ down to $\mathrm{sRep}(\mathbb{Z}_2^f)$~\cite{seo2026unified}. Thus, the resulting phase is a charge-$2e$ superconductor.

This categorical description of the FSHI-SC transition and resultant superconductor fully aligns with both the numerical results, e.g., the lack of spin Hall conductivity, and the field-theoretical analysis above.

\section{Superconductivity from Different non-Abelian FSHIs}
\paragraph{The Pfaffian and Anti-Pfaffian States} As written in the main text, the Pfaffian state is described by $\mathrm{U}(2)_{2,-4}\times\mathrm{U}(1)_8$, whose Lagrangian is given by
\begin{align*}
\mathcal{L}_\mathrm{Pf}&=-\frac{2}{4\pi}\Trace\left(c\diff c+\frac{2}{3}c^3\right)+\frac{3}{4 \pi}\left(\Trace{c}\right)\diff\left(\Trace{c}\right)\nonumber\\
&\quad+\frac{1}{2\pi}A\diff\left(\Trace{c}\right)+\frac{1}{4\pi}A\diff A,
\end{align*}
where $c$ is a $\mathrm{U}(2)$ gauge field and $A$ is the background electromagnetic gauge field. We also consider the anti-Pfaffian state is described by $\left[\mathrm{SU}(2)_{-2}\times\mathrm{U}(1)_8]/\mathbb{Z}_2\right]$, with the Lagrangian~\cite{hsin2020effective},
\begin{align*}
\mathcal{L}_\mathrm{aPf}&=\frac{2}{4\pi}\Trace\left(c\diff c+\frac{2}{3}c^3\right)-\frac{3}{4\pi}\left(\Trace c\right)\diff\left(\Trace c\right)\nonumber\\
&\quad+\frac{1}{2\pi}A\diff\left(\Trace c\right).
\end{align*}

\paragraph{The $\mathrm{aPf}\times\overline{\mathrm{aPf}}$ FSHI} The Lagrangian describing $\mathrm{aPf}\times\overline{\mathrm{aPf}}$ is
\begin{align*}
&\mathcal{L}_{\mathrm{aPf}\times\overline{\mathrm{aPf}}}=\mathcal{L}_\mathrm{aPf}+\mathcal{L}_{\overline{\mathrm{aPf}}}\nonumber\\
&=-\frac{2}{4\pi}\Trace\left(c\diff c+\frac{2}{3}c^3\right)+\frac{3}{4\pi}\left(\Trace c\right)\diff\left(\Trace c\right)\nonumber\\
&\quad-\frac{1}{2\pi}A\diff\left(\Trace c\right)\nonumber\\
&\quad+\frac{2}{4\pi}\Trace\left(c'\diff c'+\frac{2}{3}c'^3\right)-\frac{3}{4\pi}\left(\Trace c'\right)\diff\left(\Trace c'\right)\nonumber\\
&\quad+\frac{1}{2\pi}A\diff\left(\Trace c'\right).
\end{align*}
Now, we condense the bifundamental field $\Phi$ that couples to the gauge fields as $\Diff\Phi=\partial\Phi-ic\Phi-i\Phi c'$ so that $c = - c'$ is imposed in the low-energy regime. As a consequence, we get
\begin{equation*}
\mathcal{L}_{\mathrm{aPf}\times\overline{\mathrm{aPf}}}\to-\frac{2}{2\pi}A\diff(\Trace{c}).
\end{equation*}
Again, this is the effective field theory describing a charge-$2$ superconductor~\cite{hansson2004superconductors}.

\paragraph{The $\mathrm{Pf}\times\mathrm{aPf}$ FSHI} We also note that superconductivity can emerge from $\mathrm{Pf}\times\mathrm{aPf}$. The Lagrangian describing $\mathrm{Pf}\times\mathrm{aPf}$ is
\begin{align*}
&\mathcal{L}_{\mathrm{Pf}\times\mathrm{aPf}}=\mathcal{L}_\mathrm{Pf}+\mathcal{L}_{\mathrm{aPf}}\nonumber\\
&=-\frac{2}{4\pi}\Trace\left(c\diff c+\frac{2}{3}c^3\right)+\frac{3}{4\pi}\left(\Trace c\right)\diff\left(\Trace c\right)\nonumber\\
&\quad+\frac{1}{2\pi}A\diff\left(\Trace c\right)+\frac{1}{4\pi}A\diff A\nonumber\\
&\quad+\frac{2}{4\pi}\Trace\left(c'\diff c'+\frac{2}{3}c'^3\right)-\frac{3}{4\pi}\left(\Trace c'\right)\diff\left(\Trace c'\right)\nonumber\\
&\quad+\frac{1}{2\pi}A\diff\left(\Trace c'\right).
\end{align*}
Now, we condense the matter field $\Phi$ that couples to the gauge fields through $\Diff\Phi=\partial\Phi-ic\Phi+i\Phi c'$. In this case, the condensation imposes the constraint $c = c'$. As a result, we get
\begin{equation*}
\mathcal{L}_{\mathrm{Pf}\times\mathrm{aPf}}\to\frac{2}{2\pi}A\diff\left(\Trace c\right)+\frac{1}{4\pi}A\diff A.
\end{equation*}
This theory describes a charge-$2$ superconductor with chiral central charge $c_- = 1$~\cite{shi2025doping_}. Note that this is consistent with expectation because $\mathrm{Pf}$ and $\mathrm{aPf}$ have $c_- = \frac{3}{2}$ and $c_- = - \frac{1}{2}$, respectively.

\bibliography{refs.bib}